\documentclass[fleqn,usenatbib]{mnras}

\usepackage{newtxtext,newtxmath}

\usepackage[T1]{fontenc}

\DeclareRobustCommand{\VAN}[3]{#2}
\let\VANthebibliography\thebibliography
\def\thebibliography{\DeclareRobustCommand{\VAN}[3]{##3}\VANthebibliography}

\usepackage{graphicx}	
\usepackage{amsmath}	
\usepackage{subcaption}
\usepackage{bm}

\DeclareMathOperator*{\argmax}{arg\,max}

\title[TRENF for Optimal Cosmological Analysis]{Translation and Rotation Equivariant Normalizing Flow (TRENF) for Optimal Cosmological Analysis}

\author[Dai and Seljak]{
Biwei Dai,$^{1}$\thanks{E-mail: biwei@berkeley.edu}
Uro\v{s} Seljak,$^{1,2,3}$
\\
$^{1}$Department of Physics, University of California at Berkeley, Berkeley, California 94720, USA\\
$^{2}$Department of Astronomy, University of California at Berkeley, Berkeley, California 94720, USA\\
$^{3}$Lawrence Berkeley National Laboratory, Berkeley, California 94720, USA
}

\date{Accepted XXX. Received YYY; in original form ZZZ}

\pubyear{2022}

\begin{document}
\label{firstpage}
\pagerange{\pageref{firstpage}--\pageref{lastpage}}
\maketitle

\begin{abstract}
Our universe is homogeneous and isotropic, and its perturbations obey translation and rotation symmetry. In this work we develop Translation and Rotation Equivariant Normalizing Flow (TRENF), a generative Normalizing Flow (NF) model which explicitly incorporates these symmetries, defining the data likelihood via a sequence of Fourier space-based convolutions and pixel-wise nonlinear transforms.
TRENF gives direct access to the high 
dimensional data likelihood $p(x|y)$ as a function of the labels $y$, such as cosmological parameters.
In contrast to traditional analyses based on summary statistics, the NF approach has no loss of information since it preserves the full dimensionality of the data.
On Gaussian random fields, the TRENF likelihood agrees well with the analytical expression and saturates the Fisher information content in the labels $y$.
On nonlinear cosmological overdensity fields from N-body simulations, TRENF leads to significant improvements in constraining power over the standard power spectrum summary statistic. TRENF is also a generative model of the data, 
and we show that TRENF samples agree well with the N-body simulations it trained on, and that the inverse mapping of the data agrees well with a Gaussian white noise 
both visually and 
on various summary statistics: when this is perfectly achieved the resulting $p(x|y)$ likelihood analysis becomes optimal.
Finally, we develop a generalization of this model that can handle effects that break the symmetry of the data, such as the survey mask, which enables likelihood analysis on data without periodic boundaries.
\end{abstract}

\begin{keywords}
methods: data analysis -- cosmological parameters -- large-scale structure of Universe
\end{keywords}



\section{Introduction}

\label{sec:introduction}

The goal of optimal cosmological analysis is 
to extract maximal amount of information of cosmological parameters from the data. 
If the data are Gaussian distributed 
this task has a well-known solution, as one can directly evaluate the Gaussian data likelihood $p(x|y)$, where $y$ are 
cosmological parameters 
of interest. An implementation of this 
method is the  
optimal quadratic estimator \citep{Hamilton1997a, Tegmark1997a, Bond1998a}, which 
uses second-order expansion of the likelihood 
to achieve this task.
Often we must also model 
the likelihood as a function of nuisance 
parameters such as systematics, astrophysical 
sources, etc. Evaluating 
either the quadratic estimator or the likelihood 
in high dimensions is not an 
easy task in the presence of noise and survey 
mask, since it requires an inversion and a determinant of the covariance matrix, which 
for large data is prohibitively expensive. 
Often simplified versions such as the 
pseudo power spectrum analysis are adopted \citep{Hivon2002a}.
These however suffer from an absence of a simple analytic 
covariance matrix, which for optimal quadratic estimator is available in the form of the 
Fisher matrix. 

In the 
nonlinear regime, such as the large -scale structure on small scales, nonlinear gravitational 
effects in dark matter create a rapidly growing 
cascade of higher-order correlations, 
which quickly get populated at all orders. In 
this regime we often rely on N-body simulations. 
Furthermore, what we often observe are baryons, 
such as galaxy light or gas density, which have 
additional astrophysical processes that need to 
be included in the simulation. These are 
handled with nuisance parameters that try 
to parametrize the unknown physics. 

When it comes to data analysis in this 
regime the exact 
likelihood analysis is 
deemed impossible, and instead the focus 
has been on extracting information from a 
limited set of summary statistics. 
This program has numerous challenges. The first is how to  
choose the most informative summary statistics. 
While two-point function is a natural choice even 
in the nonlinear regime, adding higher -order information is less straightforward. 
Even adding the three-point function means 
adding a function of three parameters, 
which is considerably more complex to 
describe than the two-point function or its 
Fourier analog, the power spectrum. Numerous 
other ad-hoc statistics $S(x)$ have been proposed, 
from peak counts to void counts, void profiles, etc. In each case, 
one must evaluate their mean expectation as a 
function of cosmological and nuisance parameters. 
Moreover, since these are ad-hoc summary statistics their probability distribution must be evaluated numerically,  
usually by approximating the probability distribution as a multi-variate Gaussian. 
Recent developments such as Likelihood Free Inference \citep{Alsing2018a,Alsing2019a} or Simulation Based Inference \citep{Cranmer2020a} pursue this program by combining the two tasks of mean and covariance estimate into estimating the full $p(S|y)$, which can also include effects beyond the mean and covariance of the Gaussian distribution. These have the same underlying issues of summary statistics being ad-hoc and potentially sub-optimal. The 
summary statistics can also be determined 
by some information maximizing Machine Learning algorithm \citep{Ribli2019a,Charnock18}. 

An alternative approach is that of reconstruction of initial conditions \citep{Seljak2017a, Schmittfull2017a, Zhu2017a, Feng2018a, Schmidt2019a}. 
If the initial conditions are latent variables $z$, 
this approach gives either maximum posterior (MAP) of $z$ \citep{Seljak2017a, Modi2018a}, or its samples \citep{Jasche2013a, Kitaura2013a, Wang2014a}. However, 
performing the marginal integral over $z$ so 
that we are left with $p(x|y)$ has proven 
to be difficult. Samples are very 
correlated in high dimensions 
even if Hamiltonian Monte Carlo is used, and 
thousands of full N-body simulation steps may be needed between two independent samples. 
While obtaining MAP of $z$ is faster, evaluating
the marginal integral around it and obtaining 
the posterior is still expensive and can be 
suboptimal \citep{Seljak2017a}. 

In this paper we propose instead to learn 
directly the data likelihood $p(x|y)$ from 
the data simulations conditioned on $y$. Since the data is very 
high dimensional, and the simulations are 
expensive, this task has been deemed 
difficult or impossible. However, the universe
is homogeneous and isotropic on average, 
and its perturbations obey translational and 
rotational symmetry in a statistical sense. 
If these symmetries can be imposed into 
the structure of $p(x|y)$, the parametrization of the model would be greatly restricted, allowing efficient learning of $p(x|y)$. 
To see the power of symmetries we consider a simple example of modeling an N-dimensional Gaussian Random Field (GRF), where one needs $N(N+3)/2$ parameters to describe its mean and covariance matrix. However, if the GRF satisfies translation and rotation symmetry, the mean vector is reduced to a scalar, and the covariance matrix is reduced to a 1D function, i.e., the power spectrum, which can usually be parameterized by only a few parameters assuming smoothness.
This simple example shows that symmetries can greatly reduce the degrees of freedom of the model and the sample complexity. 
In this work we will use the framework of generative learning to learn the data likelihood $p(x|y)$ and 
build the symmetries into the model itself.

Latent variable generative models such as Normalizing Flows (NFs) \citep{rezende2015variational,dinh2014nice,dinh2016density,kingma2018glow,SINF}, Variational Auto-Encoders (VAEs) \citep{kingma2013auto,rezende2014stochastic} and Generative Adversarial Networks (GANs) \citep{goodfellow2014generative, radford2015unsupervised} aim to model the high-dimensional data distribution $p(x)$ by introducing a mapping from a latent variable $z$ to $x$, where $z$ is assumed to follow a given prior distribution $\pi(z)$. While all these three classes of models have been shown to produce realistic samples \citep{kingma2018glow, razavi2019generating, karras2020analyzing}, NF is the only one that allows exact density evaluation $p(x)$, and when done conditionally as $p(x|y)$ this enables a direct likelihood analysis. Another family of density estimation models is called autoregressive models \citep{germain2015made, oord2016conditional}, which decompose the high dimensional Probability Distribution Function (PDF) as the product of 1D conditional PDFs: $p(x)=\prod_{i=1}^{N} p(x_i|x_{1:i-1})$. These models require choosing a specific ordering of the pixels $x_{1:N}$ and treat the pixels differently, making it hard to enforce symmetries. For these reasons we will adopt NF as the method of choice for cosmological data analysis.

In machine learning the symmetries are often included using brute force methods such as data augmentation. This increases the amount of training data and does not reduce the complexity of the model, which must instead learn the symmetries from the data samples. There is also no guarantee that the symmetries can be perfectly learned. In such situations NFs can fail in their primary tasks, either as a realistic data generator or as a data likelihood estimator.
There are also works trying to build the symmetries into the machine learning models \citep{Cohen2016a, Weiler2019a, Worrall2017a, Wang2020a}, but these models are mostly designed for supervised tasks such as classification and high dimensional mapping, and cannot be directly used in the NF framework, which requires the learned mapping to be invertible and to have tractable Jacobian determinant. 
In this work we develop Translation and Rotation Equivariant Normalizing Flow (TRENF), which impose the symmetries explicitly into the NF model.

The novel 
developments of this paper are: 
\begin{itemize}
\item  We develop a conditional NF architecture which is translation and rotation equivariant (TRENF) for learning the likelihood $p(x|y)$ of cosmological fields. 
\item We use TRENF as a map from latent space to data space for fast generation of high dimensional simulated data conditional on cosmological parameters. Note that cosmological fields (e.g., Cosmic Microwave Background) are usually high dimensional distributions and cannot be approximated by low dimensional manifolds, and therefore modeling them with low-dimensional-manifold models like GANs could potentially introduce systematics in the samples. TRENF, on the other hand, has no dimension reduction and can sample from the full distribution $p(x|y)$ without any manifold assumption. 
\item We use TRENF as a map from the data space to the latent space, enabling visual and numerical inspection of the quality of the training: if the latent map is a perfect white noise Gaussian at the true value of $y$, then we have optimally extracted all the information from the data, encoding it into a single number $p(x|y)$. TRENF thus can identify when the 
model is incomplete, such as missing some 
systematic or modeling effect. 
Recent works on applying CNNs \citep{Ribli2019a} or novel summary statistics \citep{Cheng2020a} to extract information from the fields have shown improvements over traditional summary statistics like power spectrum, but are not 
providing any guarantees of optimality, and 
it is unclear how much information has remained 
unused. Our generative model provides a natural way to investigate this and improve upon 
these methods. 
\item We use TRENF $p(x|y)$ as a function of $y$ to directly provide uncertainty quantification via the posterior $p(y|x)=p(x|y)p(y)/p(x)$, which is the ultimate goal of a cosmological analysis. 
\item We introduce non-symmetric components into TRENF for modeling observational effects that break the symmetry of the data, such as the survey mask. 
\end{itemize}

\section{Method}

\subsection{Normalizing Flows}

Flow-based models provide a powerful framework for density estimation \citep{dinh2016density, papamakarios2017masked}
and sampling \citep{kingma2018glow}. These models map the data $x$ to latent variables $z$ through a sequence of invertible transformations $f = f_1 \circ f_2 \circ ... \circ f_n$, such that $z = f(x)$ and $z$ is mapped to a base distribution $\pi(z)$. The base distribution $\pi(z)$ is normally chosen to be a standard normal distribution, i.e. a Gaussian white noise
with zero mean and unit variance, $\pi(z)=\mathcal{N}(0,I)$. The probability density of data $x$ can be evaluated using the change of variables formula:
\begin{eqnarray}
    \label{eq:flow}
    p(x) =& \pi(f(x)) \left|\det \left(\frac{\partial f(x)}{\partial x}\right)\right| \nonumber \\
    =& \pi(f(x)) \prod_{l=1}^n \left|\det \left(\frac{\partial f_l(x)}{\partial x}\right)\right| .
\end{eqnarray}
To sample from $p(x)$, one first samples latent variable $z$ from $\pi(z)$, and then transform variable $z$ to $x$ through $x=f^{-1}(z)$. The transformation $f$ is usually parametrized with neural networks $f_{\phi}$, and the parameters $\phi$ are estimated using Maximum Likelihood Estimation (MLE):
\begin{equation}
\label{eq:MLE}
    \phi^* = \argmax_{\phi}\ \frac{1}{N}\sum_{i=1}^N\log p_{\phi}(x_i) ,
\end{equation}
where the data likelihood $p(x)$ is given by Equation \ref{eq:flow}. The MLE solution minimizes the Kullback-Leibler (KL) divergence between the model distribution $p_{\phi}(x)$ and the true data distribution. The parametrization of $f$ must satisfy the requirements that the Jacobian determinant $\det (\frac{\partial f_l(x)}{\partial x})$ is easy to compute for evaluating the density, and the transformation $f_l$ is easy to invert for efficient sampling.

\subsection{Translation and Rotation Symmetry}

It is useful to differentiate the concepts invariant and equivariant. A function $f$ is invariant if its output is unchanged when its input $x$ is transformed by a symmetry group $g$:
\begin{equation}
    f(g\cdot x) = f(x).
\end{equation}
A relevant example is the PDF of the cosmological fields, which should be invariant under translation and rotation of the fields. Similarly, a function $f$ is equivariant if its output is transformed by the same symmetry group $g$ as its input $x$:
\begin{equation}
    f(g\cdot x) = g\cdot f(x).
\end{equation}
In other words, an equivariant function commutes with the symmetry transformation. The physical laws that govern the evolution of our universe are equivariant to translation and rotation, if we view them as a mapping from the early universe to the present day. We want our NF transformation $f$ to have similar properties as the physical laws, thus to be equivariant to these symmetries. An equivariant NF $f$ also leads to invariant PDF from Equation \ref{eq:flow}.

\subsection{Translation and Rotation Equivariant Normalizing Flow (TRENF)}
\label{subsec:TRENF}

Our goal is to find a parametrization of the flow transformation $f$ such that 1) its Jacobian determinant and inverse can be efficiently calculated for likelihood evaluation (Equation \ref{eq:flow}) and sampling; 2) $f$ is equivariant to translation and rotation. The simplest form of such transformation is the Pixelwise Gaussianization (PG), which applies the same nonlinear transformation on all pixels such that the resulting one-point PDF is a standard Gaussian. This method has been used to reconstruct the primordial density fields \citep{Weinberg1992a}. However, PG is not very expressive and cannot model the correlations between different pixels. Here we 
discuss how to go beyond PG.

We observe that for any non-Gaussian PDF that is invariant to translation, one can always find a convolution kernel, such that the one-point PDF of the convolved field is non-Gaussian. This can be proven by considering the non-zero high-order cumulant of the field. Suppose the m-point cumulant is non-zero $\langle x_1 x_2 \cdots x_m\rangle_c \ne 0$ ($m>2$). We can always define a convolution kernel $T$ that is nonzero at $x_1, x_2, \cdots x_m$. The m-th 
cumulant of the one-point PDF of the convolved field contains $\langle x_1 x_2 \cdots x_m\rangle_c$ and must be nonzero for some kernel $T$.

This motivates parametrizing the flow transformation $f$ with convolutions followed by PG. Intuitively, the convolution kernels look for maximal non-Gaussianity in convolved data, which indicates non-Gaussian PDF and non-zero high-order cumulants (order above two). The PG maps the one-point PDF of the convolved data to a Gaussian and reduces the high-order cumulants. By stacking multiple such transformations, all high-order cumulants can be reduced to zero and the data distribution is mapped to a white noise Gaussian. This process can be viewed
as a generalization of SINF (sliced iterative NF) \citep{SINF} to translation equivariant data: SINF also searches for 
maximally non-Gaussian directions, followed
by PG. For translation equivariant data 
these directions are replaced with convolutions. 

Motivated by these ideas, we choose to parametrize $f$ with convolutions and pixel-wise non-linearity. Assuming periodic condition, the convolution of data $x(\bm{r})$ can be written in Fourier space as
\begin{equation}
    \int T(\bm{r}-\bm{r'})x(\bm{r'})\,d\bm{r'} = \hat{F}^{-1}\left(\tilde{T}(k)\cdot \tilde{x}(\bm{k})\right)
\end{equation}
where $T$ is the convolution kernel, $\hat{F}$ denotes Fourier transform,  $\tilde{T} = \hat{F}(T)$ and $\tilde{x} = \hat{F}(x)$ are the Fourier transform of $T$ and $x$, respectively. We require the convolution to be rotational equivariant, thus $\hat{T}$ can only depend on $k$, the amplitude of $\bm{k}$. We combine a convolution operation with a pixelwise non-linearity $\Psi$ to form the basic transformation of TRENF:
\begin{equation}
    \label{eq:TRENF}
    f = \Psi\left(\hat{F}^{-1}\tilde{T}(k)\hat{F}x\right) .
\end{equation}
Both $\tilde{T}$ and $\Psi$ are 1D functions learned from the data. We choose to parametrize $\tilde{T}(k)$ with cubic splines. $\Psi$ function is required to be monotonic and differentiable in order to sample and evaluate density from TRENF. We parametrize $\Psi$ with monotonic rational quadratic splines. We will refer to each such transformation as one layer in the rest of this paper.

Equation \ref{eq:TRENF} satisfies the two requirements we set at the beginning of this section. Firstly, its Jacobian determinant and inverse can be calculated via:
\begin{align}
    \left\vert\frac{\,df}{\,dx}\right\vert &= \left(\prod_i^{\mathrm{pixels}} \frac{\,d\Psi}{\,dx(\bm{r}_i)}\right) \cdot \left(\prod_j^{\mathrm{k\ modes}} \tilde{T}(k_j)\right) ,\\
    f^{-1} &= \hat{F}^{-1}(1/\tilde{T})\hat{F}\Psi^{-1}(x) .
\end{align}
Secondly, it can be easily verified that both the convolution and the pixel-wise non-linearity are translational and rotational equivariant, so Equation \ref{eq:TRENF} also satisfies the symmetry requirement. To improve the expressivity of the model, one can stack multiple such transformations and form a deep NF model. 

The architecture of TRENF is similar to a Convolutional Neural Network (CNN): both of them are composed of convolutions followed by non-linearities. While CNNs normally perform convolution in real space,
this becomes too expensive for long 
range correlations typical of 
cosmology data. 
TRENF compute the convolution in Fourier space, making it possible to easily calculate its inverse and Jacobian determinant. The Fourier space also allows us to easily enforce rotational symmetry, and parametrize arbitrarily large kernels. In the NF framework we keep the dimensionality of the data, unlike CNNs which usually change the number of channels and side lengths. Another difference between TRENF and CNN is that the non-linearity is learnable in TRENF, which in CNN it is normally chosen to be a fixed function like ReLU. This extra degree of freedom increases the flexibility of TRENF.

\subsection{Conditional TRENF}

To learn the model dependence on cosmology, baryonic physics, and other nuisance parameters, we build conditional TRENF where the model parameters are functions of conditional variables $y$. Specifically, we train a hyper neural network $g$ to learn the conditional relation
 $   \phi=g(y)$, 
where $\phi$ consists of the spline parameters of all the kernel $\tilde{T}$ and non-linearity $\Psi$.
The total number of TRENF parameters $\phi$ is usually of order $\mathcal{O}(100)$, and they can be easily predicted by a single fully-connected hyper network. An 
alternative is to interpolate between 
different $y$ using a Gaussian Process.


\subsection{Training}
\label{subsec:training}
We explore two kinds of training losses in this work: generative loss and discriminative loss. In the generative loss we minimize the negative log-likelihood, which is the standard loss function of NF (Equation \ref{eq:MLE} with conditional variable y):
\begin{equation}
\label{eq:Lg}
\mathcal{L}_{\mathrm{g}} = -\frac{1}{N}\sum_{i=1}^N\log p(x_i|y_i) .
\end{equation}
The generative training is suitable for sampling (Section \ref{subsec:sample}) and conditional density estimation. 

For posterior analysis in Section \ref{subsec:posterior_simulation}, our task is to obtain the most accurate posterior distribution, rather than the likelihood function. It has also been shown that discriminative learning with objective $p(y|x)$ generally has a lower asymptotic error on classification tasks than generative learning with objective $p(x|y)$ \citep{ng2002discriminative}. Therefore, we adopt a two-stage training strategy, where we firstly train TRENF with a generative loss, and then we switch to the discriminative loss ($-\log p(y|x)$) to improve the accuracy of the posteriors. The first stage (generative learning) can be viewed as an initialization (warm startup) of the discriminative learning and speeds up the training process.
In the second stage, the calculation of $\log p(y|x)$ involves computing the evidence $p(x)$, which is estimated using Importance Sampling (IS):
\begin{align}
\label{eq:Ld}
&    \mathcal{L}_{\mathrm{d}} = -\frac{1}{N}\sum_{i=1}^N\log p(y_i|x_i) \nonumber \\
&    = -\frac{1}{N}\sum_{i=1}^N \left[ \log p(x_i|y_i)+ \log p(y_i) - \log \left(\int p(x_i|y) p(y) dy\right)
    \right] \nonumber\\
    &\approx  -\frac{1}{N}\sum_{i=1}^N \left[\vphantom{\frac{1}{M}\sum_{y_j \sim q(y|x_i)}^M} \log p(x_i|y_i) + \log p(y_i)\right.\nonumber\\
    & -\left. \log \left(\frac{1}{M}\sum_{y_j \sim q(y|x_i)}^M \frac{p(x_i|y_j)p(y_j)}{q(y_j|x_i)}\right) \right] 
    .
\end{align}
During training, for each training data $x_i$, we first find the MAP solution $y_{i,\mathrm{MAP}}=\argmax p(x_i|y)p(y)$ using ADAM optimizer, and then the IS distribution $q(y|x_i)$ is defined as a Gaussian centered at $y_{i,\mathrm{MAP}}$ with a fixed covariance matrix. The parameters of the covariance matrix are fitted to the posterior distribution of the first stage training. 
Note that if we skip the first stage training and directly train TRENF with Equation \ref{eq:Ld}, the Gaussian $q(y|x_i)$ is normally a poor approximation to the true posterior, and the optimization is difficult to converge due to inaccurate estimation of $p(x)$.
The number of importance sampling points $M$ is a hyperparameter, and we use $M=20$ for the datasets we considered in this paper. 

\begin{figure*}
     \centering
     \begin{subfigure}[]{0.49\linewidth}
         \centering \includegraphics[width=\linewidth]{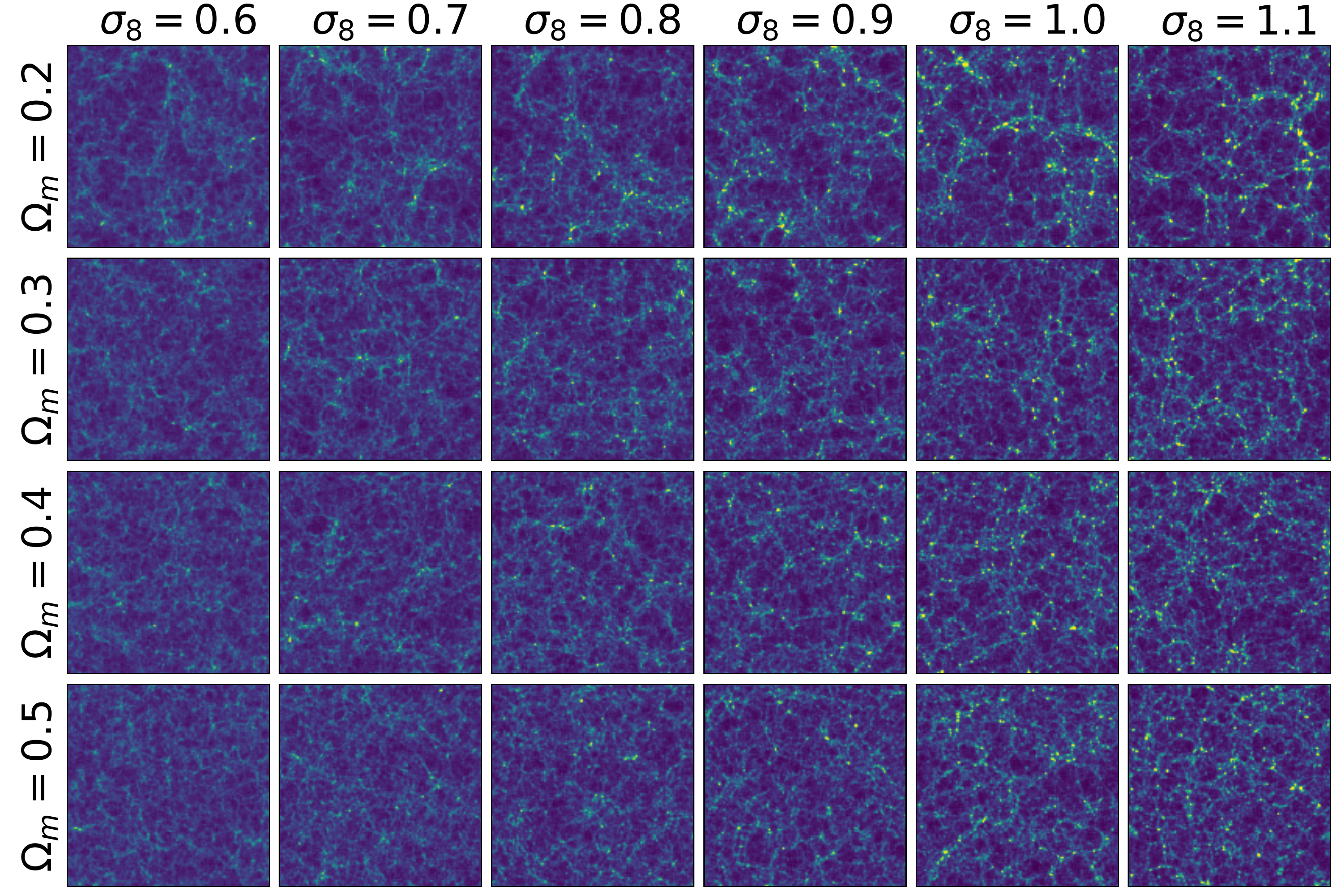}
     \end{subfigure}
     \hfill
     \begin{subfigure}[]{0.49\linewidth}
         \centering \includegraphics[width=\linewidth]{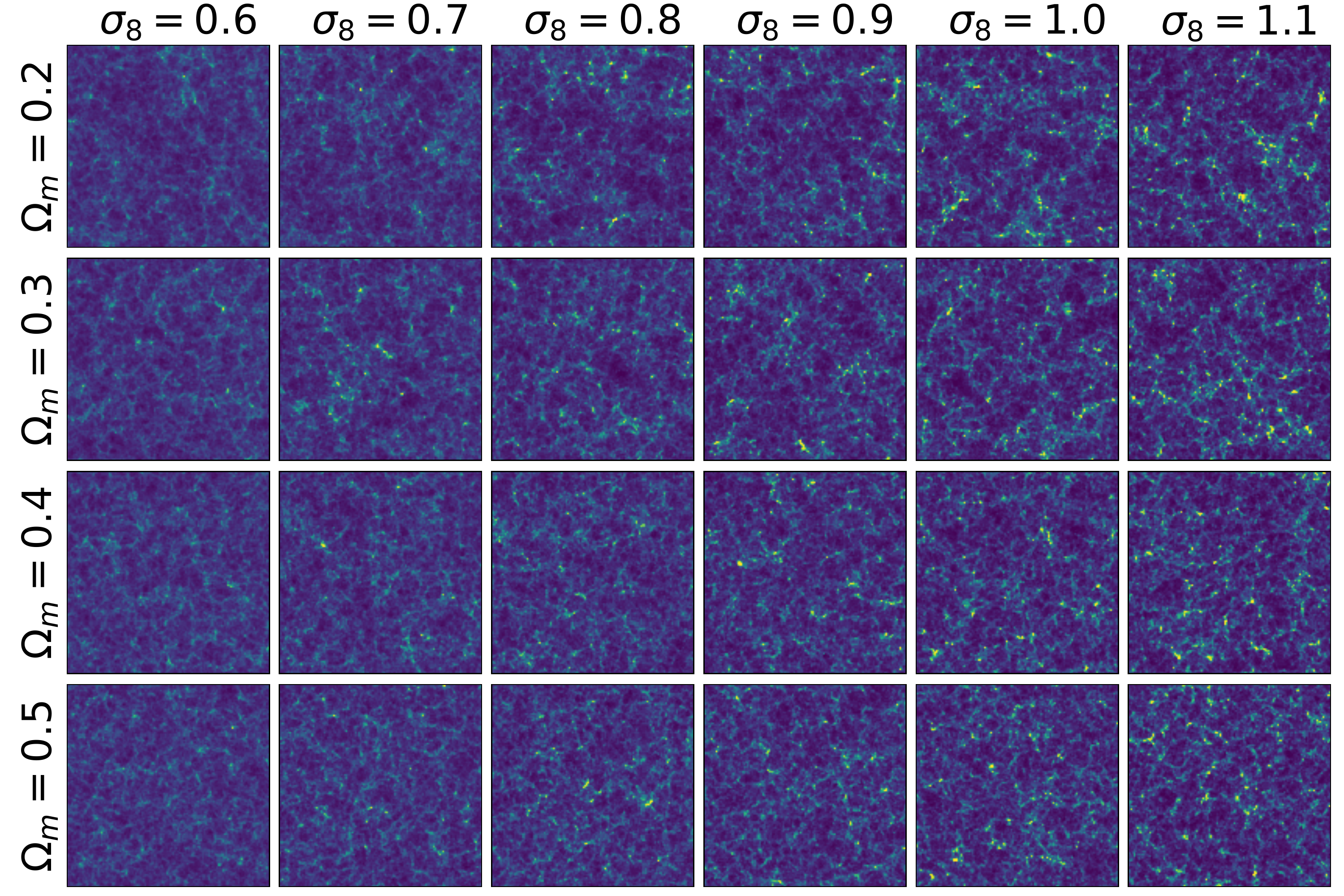}
     \end{subfigure}
    \caption{Test data (left panel) and uncurated TRENF samples (right panel) as a function of conditional variables $\Omega_m$ and $\sigma_8$.}
    \label{fig:sample}
\end{figure*}

\subsection{Modeling Effects that Break the Data Symmetry}

\label{subsec:mask}

The physical fields satisfy translation and rotation symmetries, but our observed data usually do not. There are several observational effects, such as the survey mask and foregrounds, that could break the symmetry of the data. To model these effects we need to introduce non-equivariant components into our model. The idea is that the non-equivariant component will model these observational effects that break the symmetries, while TRENF takes care of the physical process that obeys the symmetries. 

As data preprocessing, we first sample Gaussian random noise to the missing pixels $x_{\mathrm{mask}}$ so that the data have rectangular shapes. We then add an affine coupling layer \citep{dinh2016density} which applies affine transforms on $x_{\mathrm{mask}}$, conditional on the neighboring pixels $x_{\mathrm{neighbor}}$ and conditional variables $y$:
\begin{equation}
\label{eq:affine}
    z = x_{\mathrm{mask}} \odot \exp(s(x_{\mathrm{neighbor}}, y)) + t(x_{\mathrm{neighbor}}, y) ,
\end{equation}
where $\odot$ denotes element-wise product, and $s$ and $t$ are functions modeled by neural networks. The other pixels are left unchanged in this layer. This step can be effectively seen as inpainting, even though we do not explicitly train the layer to recover the missing pixels, but rather train the whole model using NF objectives.

After the affine coupling layer, we add convolutions and pixel-wise non-linearities similar to Equation \ref{eq:TRENF} to map the data to a Gaussian. Here we introduce non-equivariant components into these transforms to model effects like non-periodic boundaries and position-dependent noise.
Since these effects are usually position-dependent, in this work we choose to introduce position dependence on the non-linearity $\Psi(x) = \Psi_{\bm{r}}(x)$. Specifically, we train two separate hyper networks $g_{\tilde{T}}$ and $g_{\Psi}$. $g_{\tilde{T}}$ models the dependence of convolution kernel parameters $\phi_{\tilde{T}}$ on conditional variable $y$:
\begin{equation}
\label{eq:phi_T}
    \phi_{\tilde{T}} = g_{\tilde{T}}(y),
\end{equation}
while $g_{\Psi}$ models the conditional dependence of non-linearity parameters $\phi_{\Psi}$ on position $\bm{r}$ and $y$: 
\begin{equation}
\label{eq:phi_Psi}
    \phi_{\Psi} = g_{\Psi}(\bm{r}, y).
\end{equation}



\section{Results: generative samples in data space and data representation in latent space}
\label{sec:sample_latent}

{\bf Dataset:}
The dataset we will use throughout the paper is 2D projections of matter overdensity fields. This example is most relevant for weak lensing applications, which are similar projections of matter density field along the line of sight. Here, for the initial analysis, we want to have the data to be periodic, so that translation and rotation symmetry is not broken. In section \ref{sec:mask} we will generalize it to non-periodic boundary. 

The matter overdensity fields are generated by N-body solver FastPM \citep{Feng2016a}. We uniformly sample $\Omega_m$ and $\sigma_8$ from the prior $\Omega_m \in [0.2, 0.5]$ and $\sigma_8 \in [0.5, 1.1]$, and fix the other cosmological parameters to Planck 2015 \citep{Planck2016a}. $\Omega_m$ and $\sigma_8$ are the conditional variable $y$ in this study. Each simulation is run in a 512 $h^{-1}\mathrm{Mpc}$ box with 10 time steps using $128^3$ particles. The matter overdensity field at redshift $0$ is measured on a $128^3$ mesh. Then we divide the box into four slices (128 $h^{-1}\mathrm{Mpc}$ thick) along z-axis, and project each slice along z-axis to get four $128^2$ matter overdensity fields. 

{\bf TRENF:}
We build a TRENF model consisting of 5 transformation blocks (Equation \ref{eq:TRENF}), with $8$ spline points in $\tilde{T}$ and $\Psi$ in each block. The hyper network is chosen to be a multilayer perceptron with $2$ hidden layers and $512$ neurons in each hidden layer. Since our input data $1+\delta$ is non-negative, the data preprocessing is performed by first removing the $[0,+\infty)$ boundary with an inverse softplus transform 
\begin{equation}
    \mathrm{invsoftplus}(1+\delta) = \log (e^{1+\delta}+\epsilon-1) ,
\end{equation}
followed by a normalization layer $\frac{x-\mu}{\sigma}$ to scale the data to zero mean and unity variance. Here $\mu$ and $\sigma$ are both scalars that are independent of conditional variable $y$.
The generative loss function (Equation \ref{eq:Lg}) is used to optimize the model in the first stage.

\subsection{Generative samples in data space}
\label{subsec:sample}

\begin{figure}
     \centering
     \begin{subfigure}[]{0.49\linewidth}
         \centering \includegraphics[width=\linewidth]{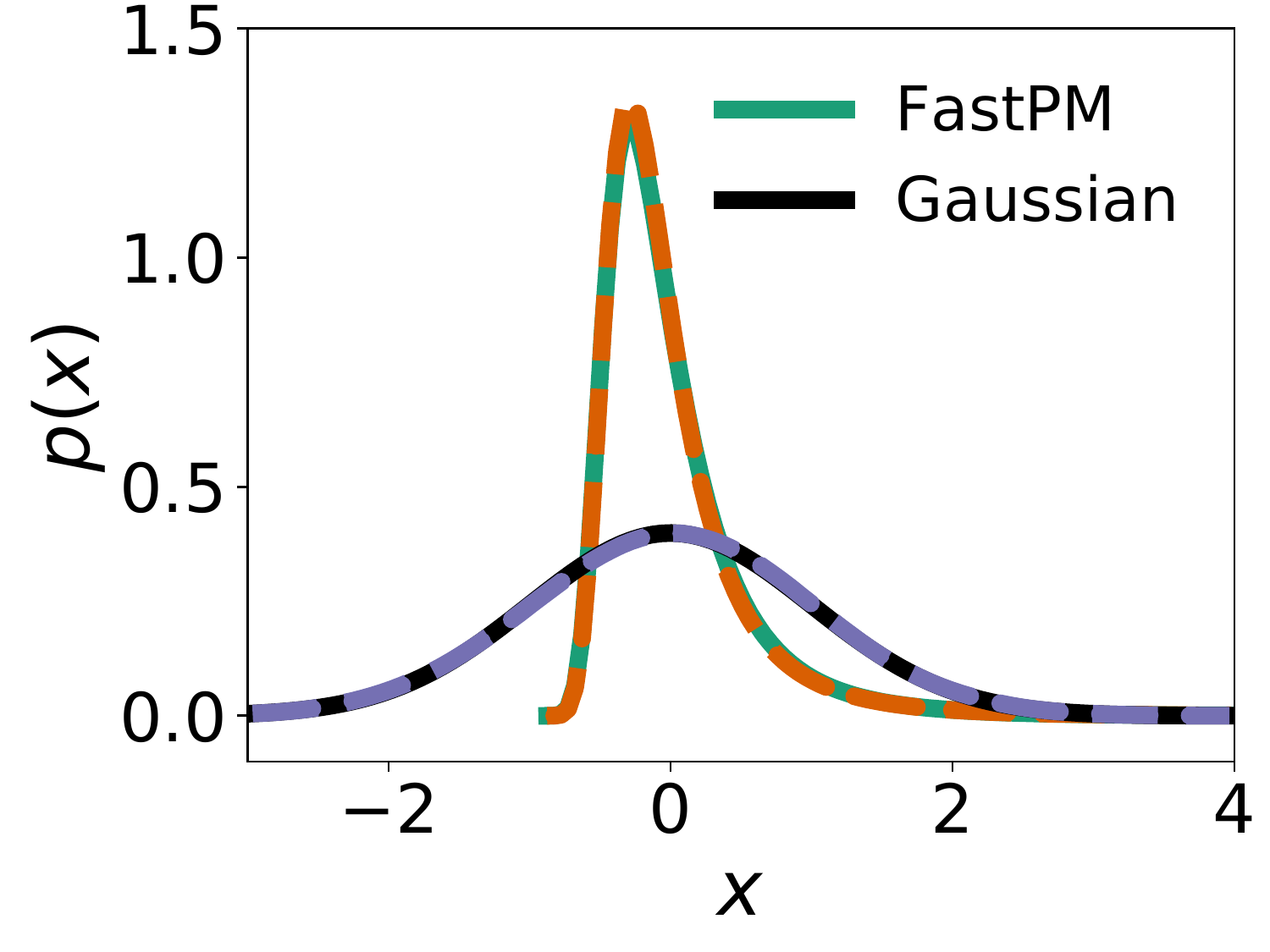}
     \end{subfigure}
     \hfill
     \begin{subfigure}[]{0.49\linewidth}
         \centering \includegraphics[width=\linewidth]{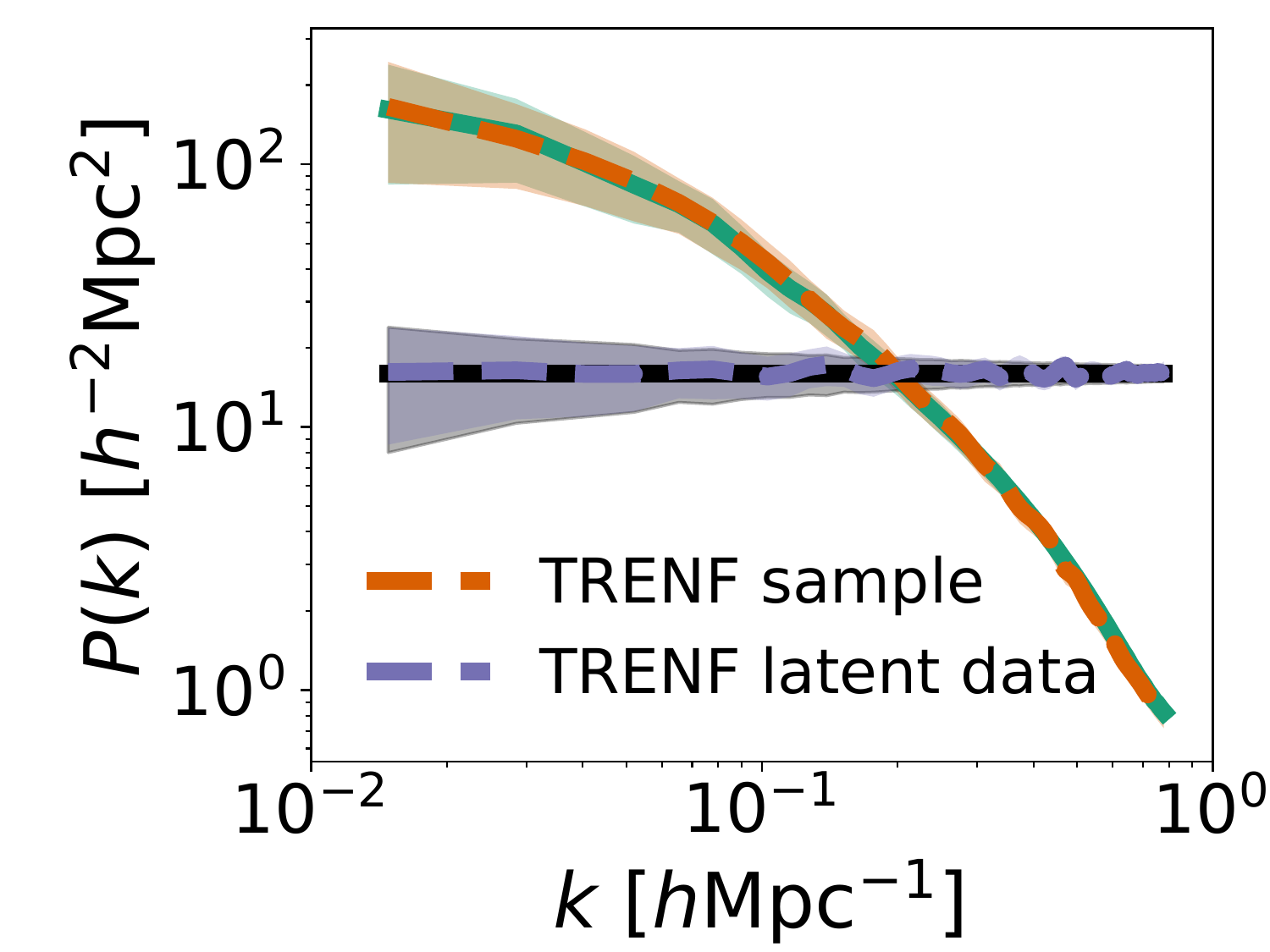}
     \end{subfigure}
     \hfill
     \begin{subfigure}[]{0.49\linewidth}
         \centering \includegraphics[width=\linewidth]{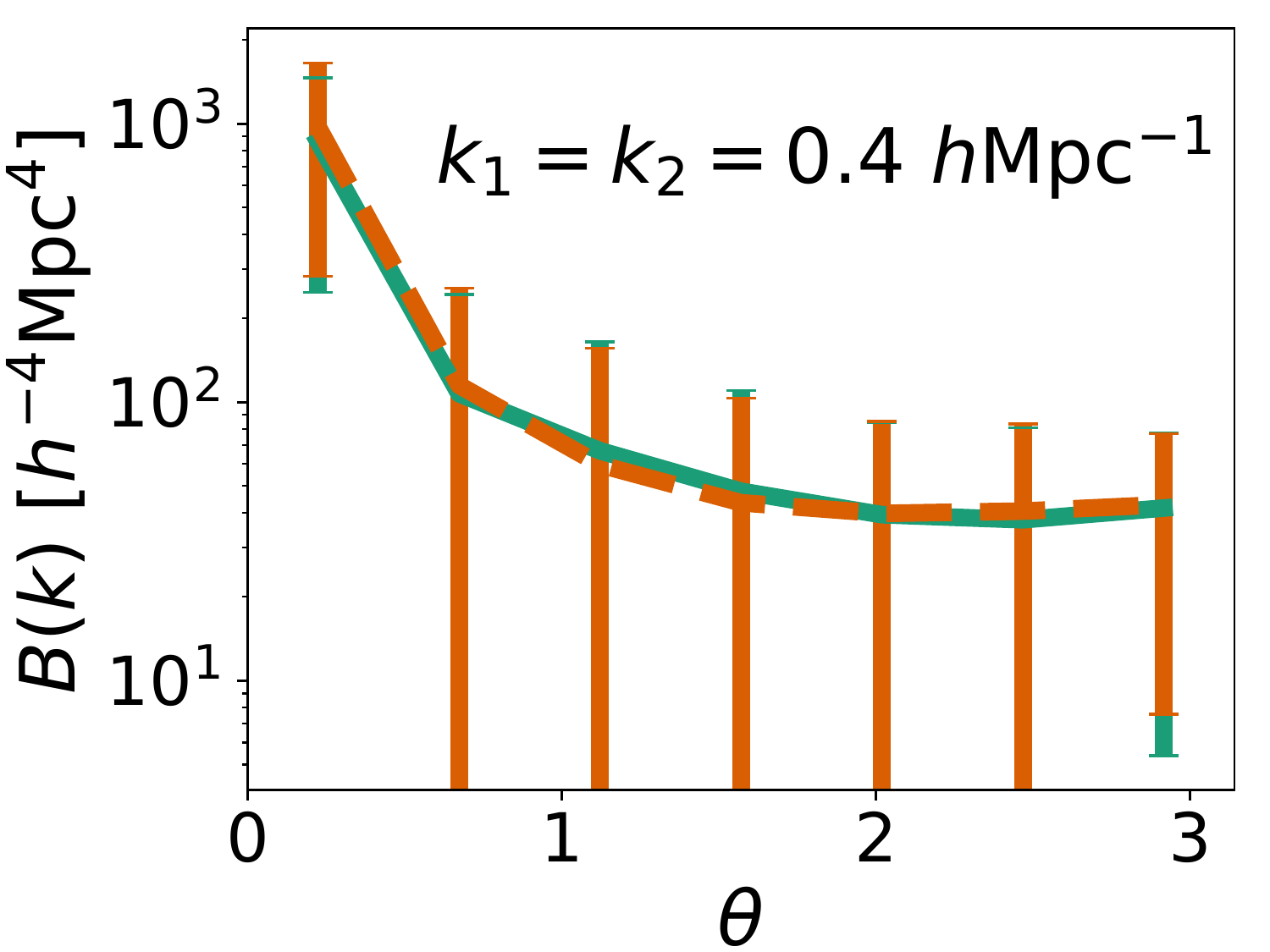}
     \end{subfigure}
     \hfill
     \begin{subfigure}[]{0.49\linewidth}
         \centering \includegraphics[width=\linewidth]{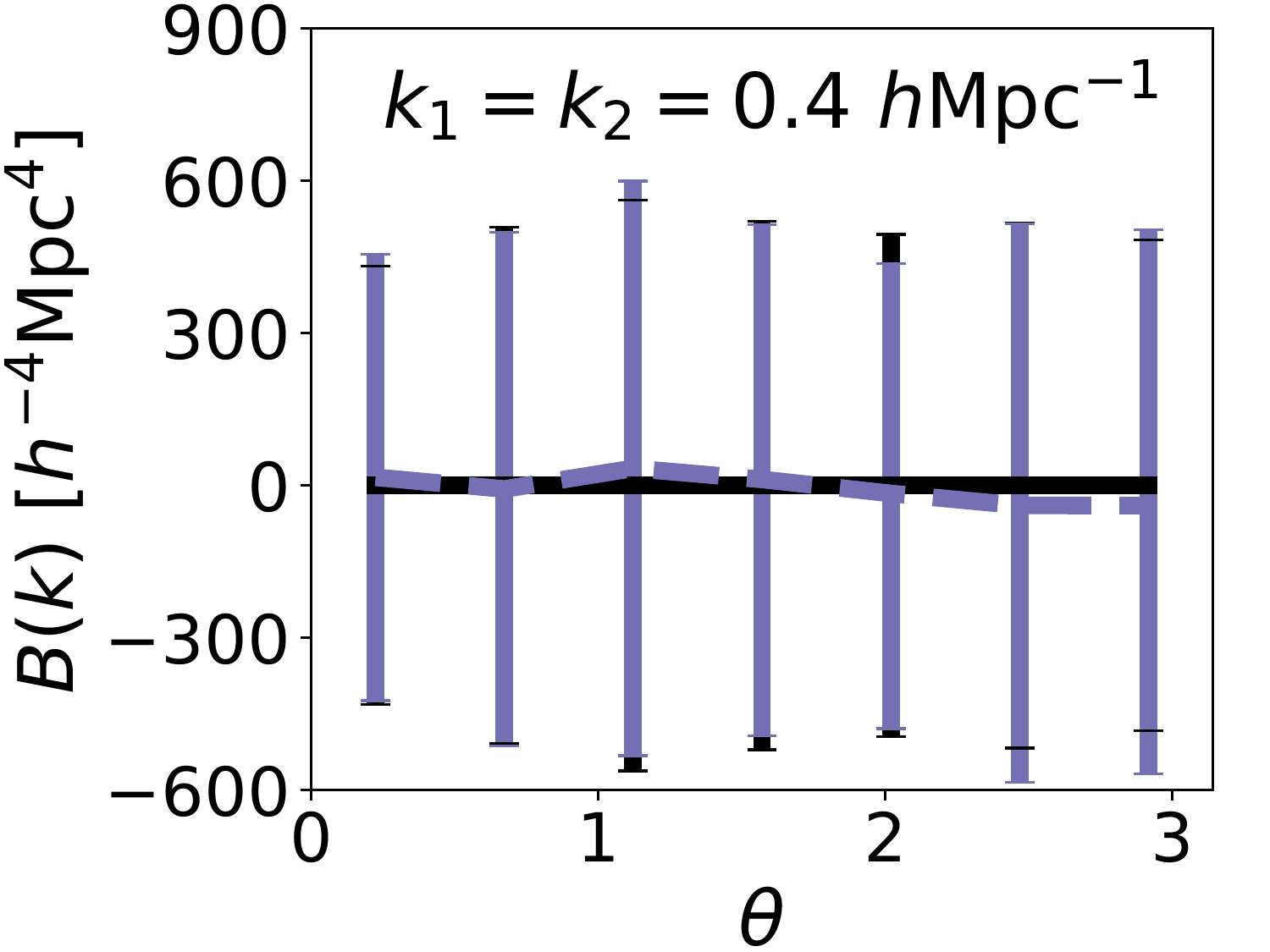}
     \end{subfigure}
    \caption{The pixel probability distribution function (upper left), power spectrum (upper right), and bispectrum (lower panels) of FastPM, TRENF samples, and FastPM data in TRENF latent space. All the results are measured over 10000 samples. The shadowed regions in the power spectrum plot and the error bars in the bispectrum plot indicate $16\%$ and $84\%$ of the distribution. The samples of TRENF agree well with FastPM on these summary statistics. In TRENF latent space FastPM data is consistent with Gaussian white noise.}
    \label{fig:summary_statistics}
\end{figure}

Once we have trained the NF we can draw a vector $z$ from a white noise distribution and map it into the data space via $x=f_y^{-1}(z)$. Figure \ref{fig:sample} shows the resulting maps sampled from the trained TRENF, comparing them to the test data. The training and sampling are conditioned on cosmological parameters $\Omega_m$ and $\sigma_8$. We see that the samples have a similar structure as the test data, and reproduce the nonlinear evolution of structure with $\sigma_8$ (structure becoming more nonlinear with $\sigma_8$), and voids becoming smaller
with $\Omega_m$. 

In Figure \ref{fig:summary_statistics} we show various statistics 
run on test data and on TRENF samples. We compare
them in terms of the power spectrum, one-point 
distribution function at the pixel scale, 
and the bispectrum. In all cases the agreement 
is nearly perfect, suggesting that TRENF samples are not only visually correct but also 
reproduce the low and high order statistics. 

TRENF takes about $0.4$ second to generate 100 images on a Tesla V100 GPU. The simulations 
we trained on are computationally cheap, as they are generated with fast Particle-Mesh simulations FastPM with only 10 time steps, 
so the computing time is
about $40$ seconds on a CPU. One could also train TRENF on output maps obtained from full N-body simulations or hydrodynamical simulations, and the computational gain in sampling time of a simulation image would be more significant. This shows the promise
of TRENF as a fast and realistic generative 
model for cosmological data such as galaxy 
images, weak lensing maps, Sunyaev-Zeldovich
maps (thermal and kinetic), etc. 

\subsection{Data representation in latent space}

\begin{figure}
     \centering
     \begin{subfigure}[]{0.325\linewidth}
         \centering \includegraphics[width=\linewidth]{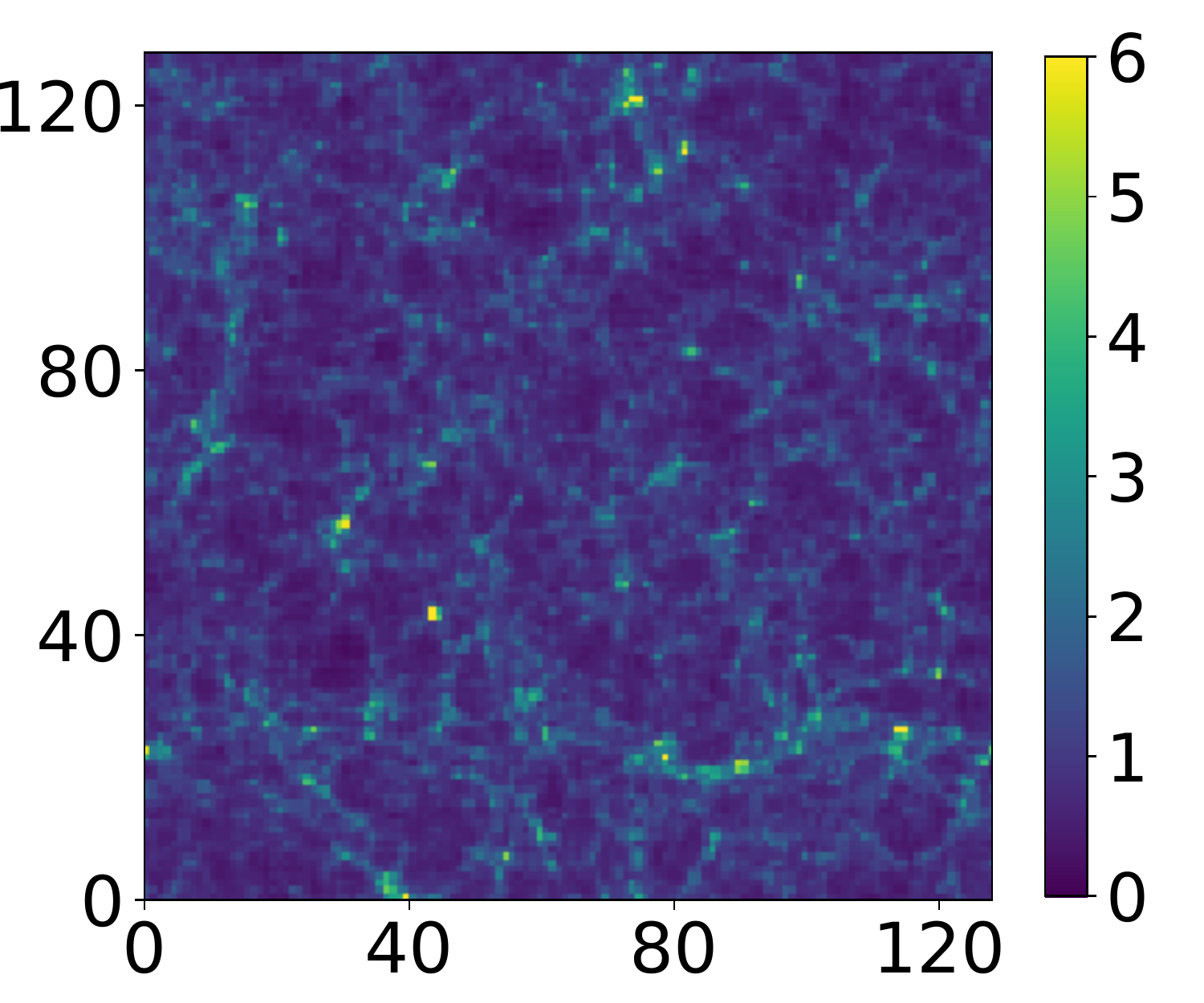}
     \end{subfigure}
     \hfill
     \begin{subfigure}[]{0.325\linewidth}
         \centering \includegraphics[width=\linewidth]{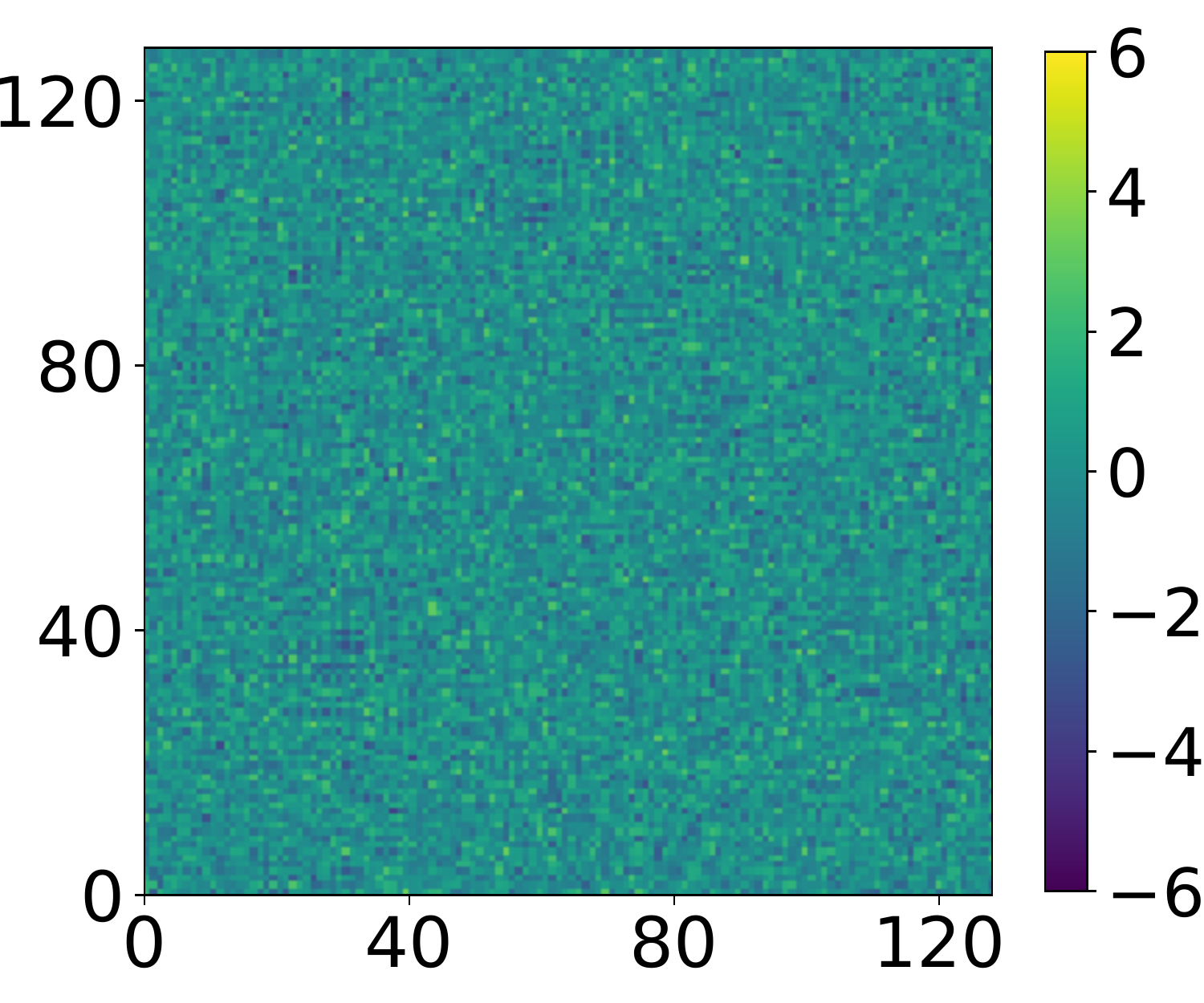}
     \end{subfigure}
     \hfill
     \begin{subfigure}[]{0.325\linewidth}
         \centering \includegraphics[width=\linewidth]{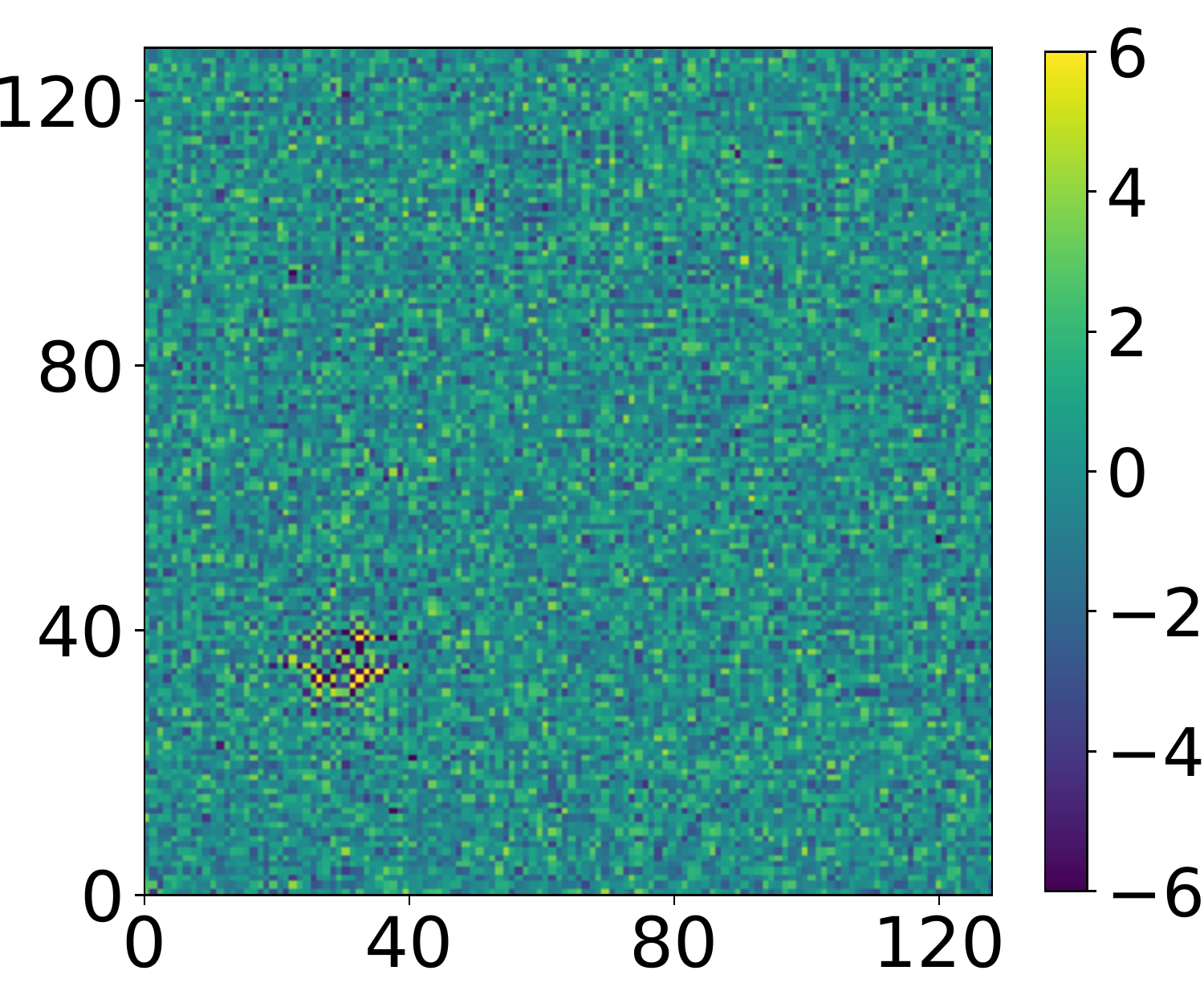}
     \end{subfigure}
    \caption{Test data (left panel), latent data transformed with correct cosmology ($\Omega_m=0.309$, $\sigma_8=0.816$, middle panel), and latent data transformed with incorrect cosmology ($\Omega_m=0.4$, $\sigma_8=0.5$, right panel).}
    \label{fig:latent}
\end{figure}

The training of TRENF achieves its goal of optimal likelihood if it maps the data into the target distribution, for which we use a Gaussian white noise distribution. 
To test this we compare the test data mapped into the latent space with the standard Gaussian distribution. 
In Figure \ref{fig:latent} we show the visualization of the latent data transformed with the correct conditional variable $y$ and incorrect $y$. We can see that when the correct $y$ is used the latent space is visually indistinguishable from Gaussian white noise. We show their one-point PDFs, power spectra and bispectra in Figure \ref{fig:summary_statistics}. 
On all these summary statistics the latent data are consistent with the standard Gaussian.
This is very encouraging: 
if the latent data distribution is a perfect Gaussian white noise, then we have achieved optimal NF, and the resulting $p(x|y)$ contains all the information of the data $x$. 
In contrast, when we use an incorrect $y$ the map is no longer white Gaussian. For example, we see strange patterns in the latent map at the position of a large void in the data space (Figure \ref{fig:latent}). 


\section{Results: likelihood and posterior analysis}

\label{sec:posterior}

From the perspective of optimal cosmological 
analysis, 
the most powerful component of NFs is their 
ability to provide conditional density or 
likelihood $p(x|y)$. If the likelihood 
is extracted 
optimally then we can achieve optimal 
cosmological analysis. To establish the ability of 
TRENF to extract the likelihood  we turn 
first to a Gaussian Random Field (GRF) example, where the 
information content of the data and the 
likelihood of the data are both known analytically.

\subsection{Gaussian Random Fields}

Similar to the matter overdensity map, we generate GRFs $\delta(r)$ in 512 $h^{-1}\mathrm{Mpc}$ boxes with $128^2$ resolution. The halofit power spectrum \citep{takahashi2012a} at redshift $0$ is used to generate the data with cosmological parameters $\Omega_m$ and $\sigma_8$ uniformly sampled from the same range $\Omega_m \in [0.2, 0.5]$ and $\sigma_8 \in [0.5, 1.1]$ \footnote{For simplicity, we assume the numerical value of the 2D GRF power spectrum is equal to the 3D halofit matter power spectrum of the same $k$ amplitude, i.e., $P_{\mathrm{GRF}}(k) = P_{\mathrm{halofit}}(k)h\mathrm{Mpc}^{-1}$.}. We build a one-layer TRENF model with $16$ spline points in $\tilde{T}$ and $8$ spline points in $\Psi$. The architecture of the hyper network and the training process are the same as Section \ref{sec:sample_latent}.

\begin{figure}
     \centering
     \begin{subfigure}[]{0.49\linewidth}
         \centering \includegraphics[width=\linewidth]{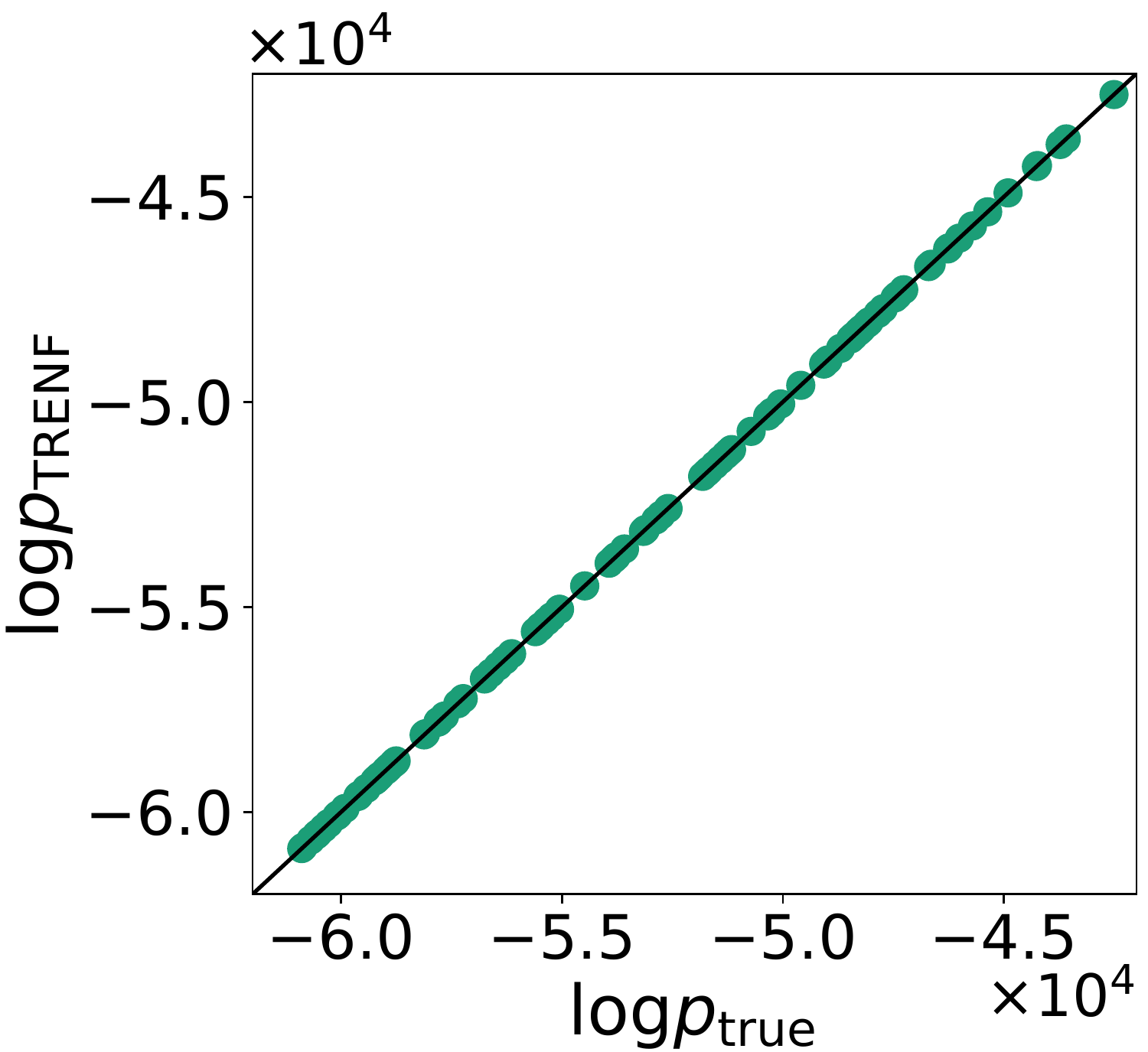}
     \end{subfigure}
     \hfill
     \begin{subfigure}[]{0.49\linewidth}
         \centering \includegraphics[width=\linewidth]{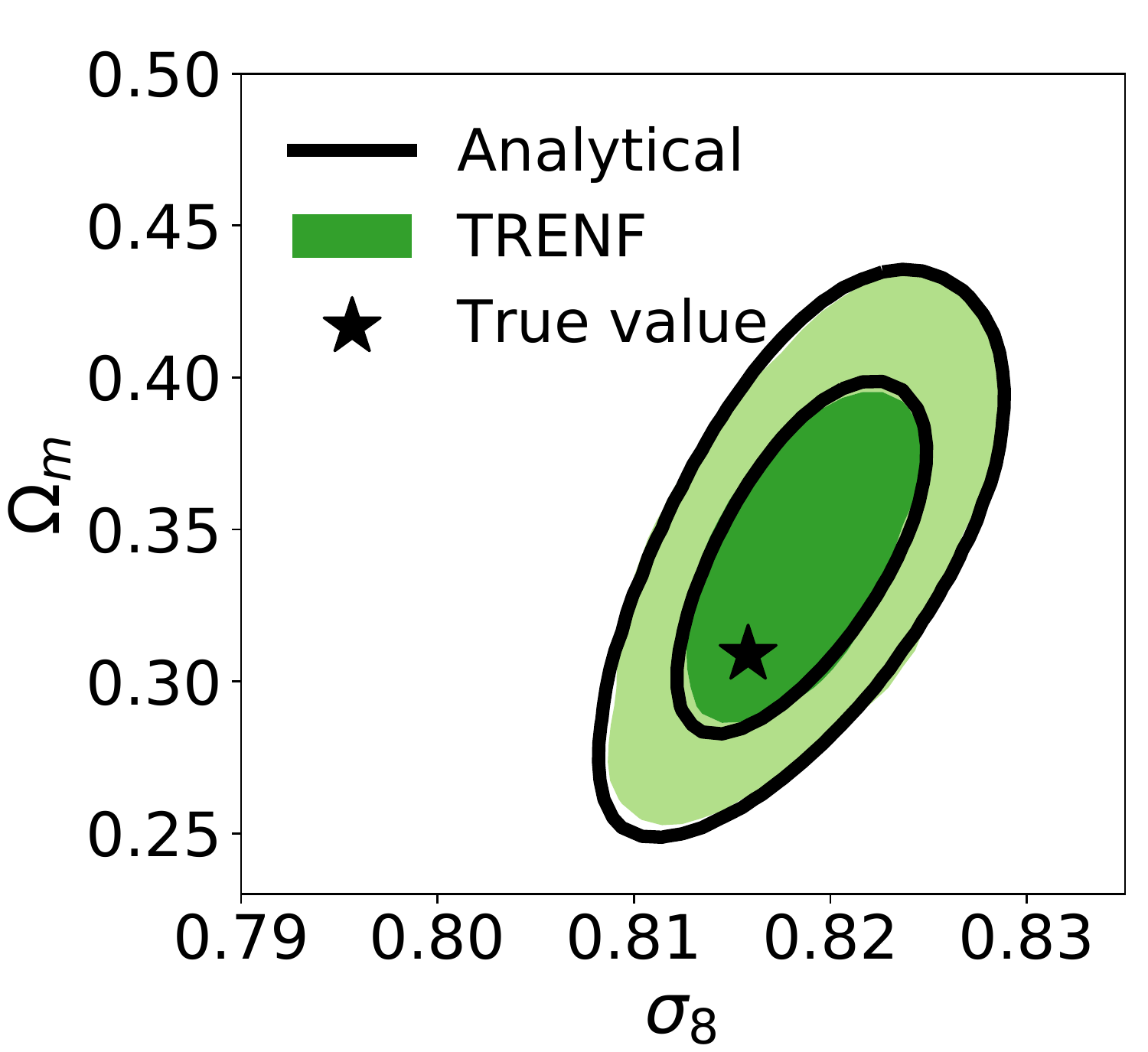}
     \end{subfigure}
    \caption{The comparison of log-likelihood (left panel) and posterior (right panel) between TRENF and analytical expression for Gaussian random fields from test set.}
    \label{fig:Gaussian}
\end{figure}

\begin{figure}
     \centering
     \begin{subfigure}[]{0.9\linewidth}
         \centering \includegraphics[width=\linewidth]{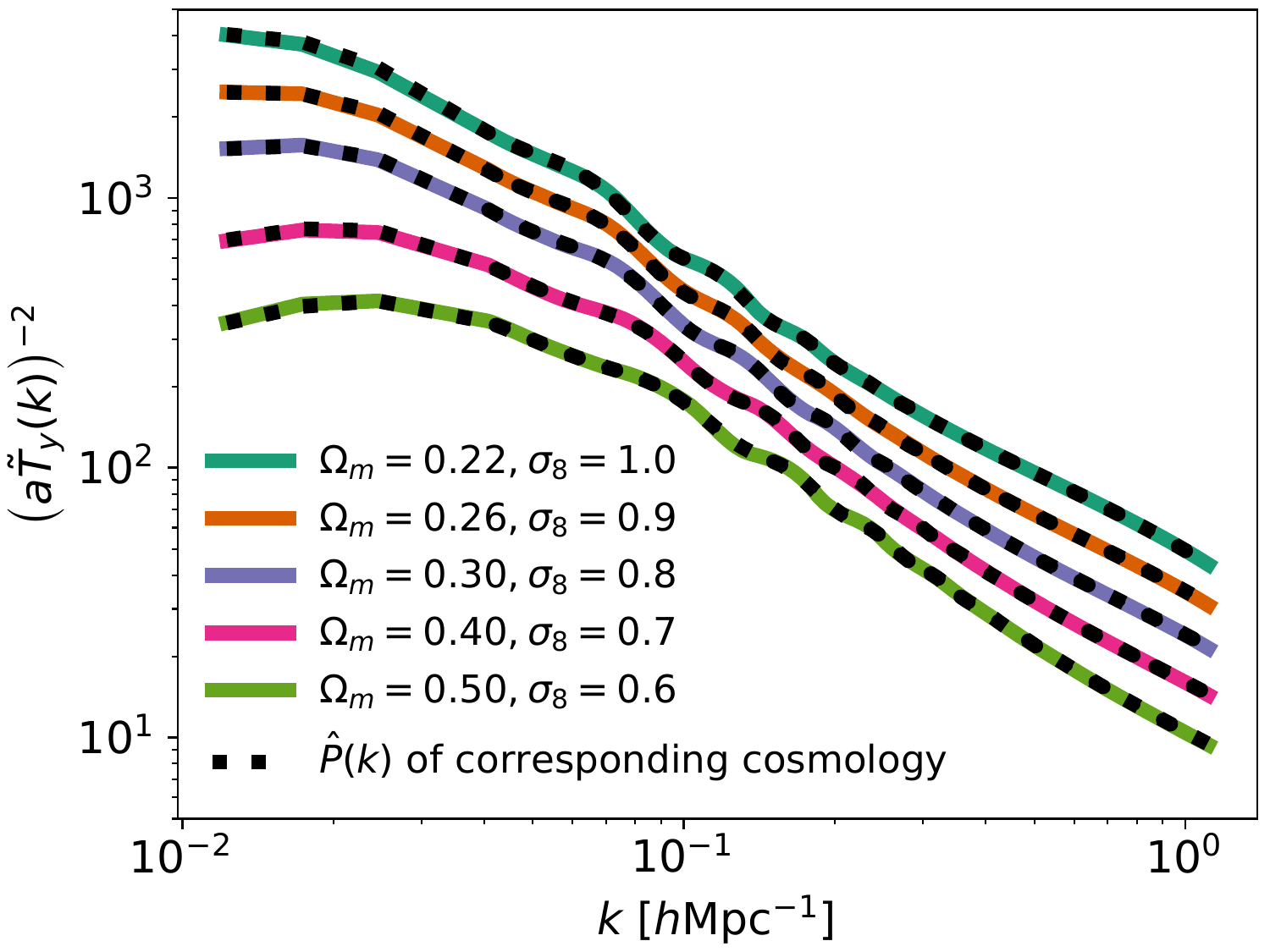}
     \end{subfigure}
     \begin{subfigure}[]{0.91\linewidth}
         \centering \includegraphics[width=\linewidth]{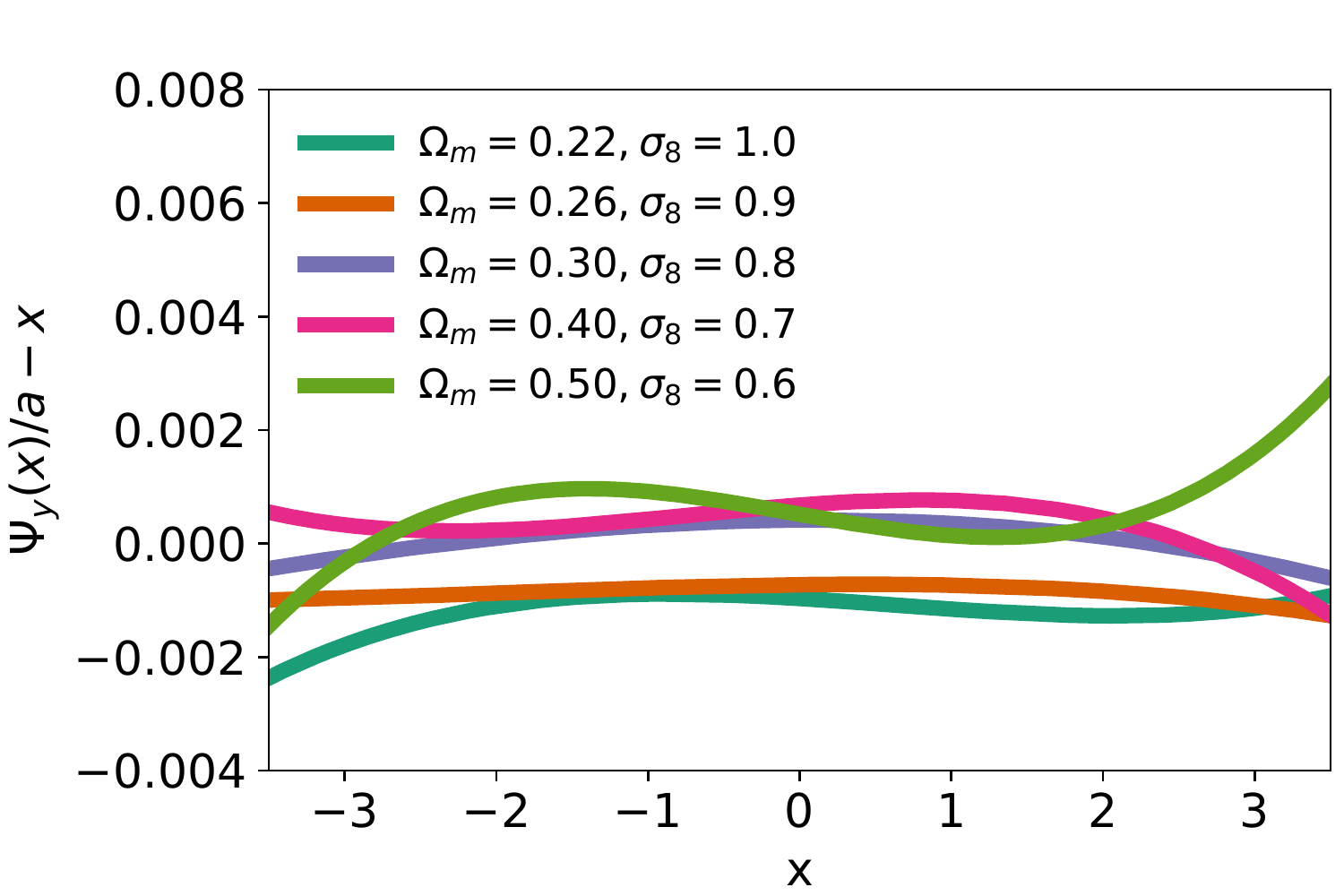}
     \end{subfigure}
    \caption{The comparison of the learned convolutional kernel $\tilde{T}(k)$ (top panel) and non-linearity $\Psi_y(x)$ (bottom panel) with the optimal solution (Equation \ref{eq:GRF_T} and \ref{eq:GRF_Psi}) for GRFs.}
    \label{fig:Gaussian2}
\end{figure}

We compare the learned likelihood from TRENF with the true analytical likelihood
\begin{equation}
    \log L_{G} = -\sum_{k}\frac{|\tilde{\delta}(k)|^2}{2\hat{P}_y(k)} - \frac{128^2}{2}\log(2\pi) - \frac{1}{2}\sum_k \log \hat{P}_y(k),
\end{equation}
where $\hat{P}_y(k) = \frac{N^2}{L^2}P_y(k)$ is the covariance of the dimensionless $\tilde{\delta}(k)$, with $L=512h^{-1}\mathrm{Mpc}$ denoting the box size and $N=128$ denoting the mesh size.
In Figure \ref{fig:Gaussian} we show the likelihood comparison on test data with random cosmology. We also show the posterior distribution from TRENF on test data and compare it with the posterior from the analytical likelihoods. 
The TRENF likelihood and posterior agree very well with the true answer, suggesting that TRENF is able to extract all the information from the GRFs.

In fact, the optimal solution of TRENF can be written down analytically for the GRFs:
\begin{align}
    \tilde{T}_y(k) &= \frac{1}{a\sqrt{\hat{P}_y(k)}}\label{eq:GRF_T} ,\\
    \Psi_y(x) &= ax\label{eq:GRF_Psi} ,
\end{align}
where $a\neq 0$ is a free coefficient that represents the degeneracy between $\tilde{T}_y(k)$ and $\Psi_y(x)$, and which 
cancels out in Jacobian determinant. This allows us to explicitly check whether TRENF has found the optimal solution. In Figure \ref{fig:Gaussian2} we show the learned $(a\tilde{T}_y(k))^{-2}$ and compare it with the scaled power spectrum $\hat{P}_y(k) = \frac{N^2}{L^2}P_y(k)$ for different cosmologies, where the coefficient $a$ is measured by fitting a linear relation between $x$ and $\Psi_y(x)$. We also present the difference between the learned $\Psi_y(x)$ and the true solution,  $\Psi_y(x)/a - x$: the two agree with each other to about one part in a thousand across the entire range of $x$. These results demonstrate that the training of TRENF converges to the correct solution (Equation \ref{eq:GRF_T} and \ref{eq:GRF_Psi}).

\subsection{Matter Overdensity Fields}

\label{subsec:posterior_simulation}

\begin{figure}
    \centering
    \includegraphics[width=\linewidth]{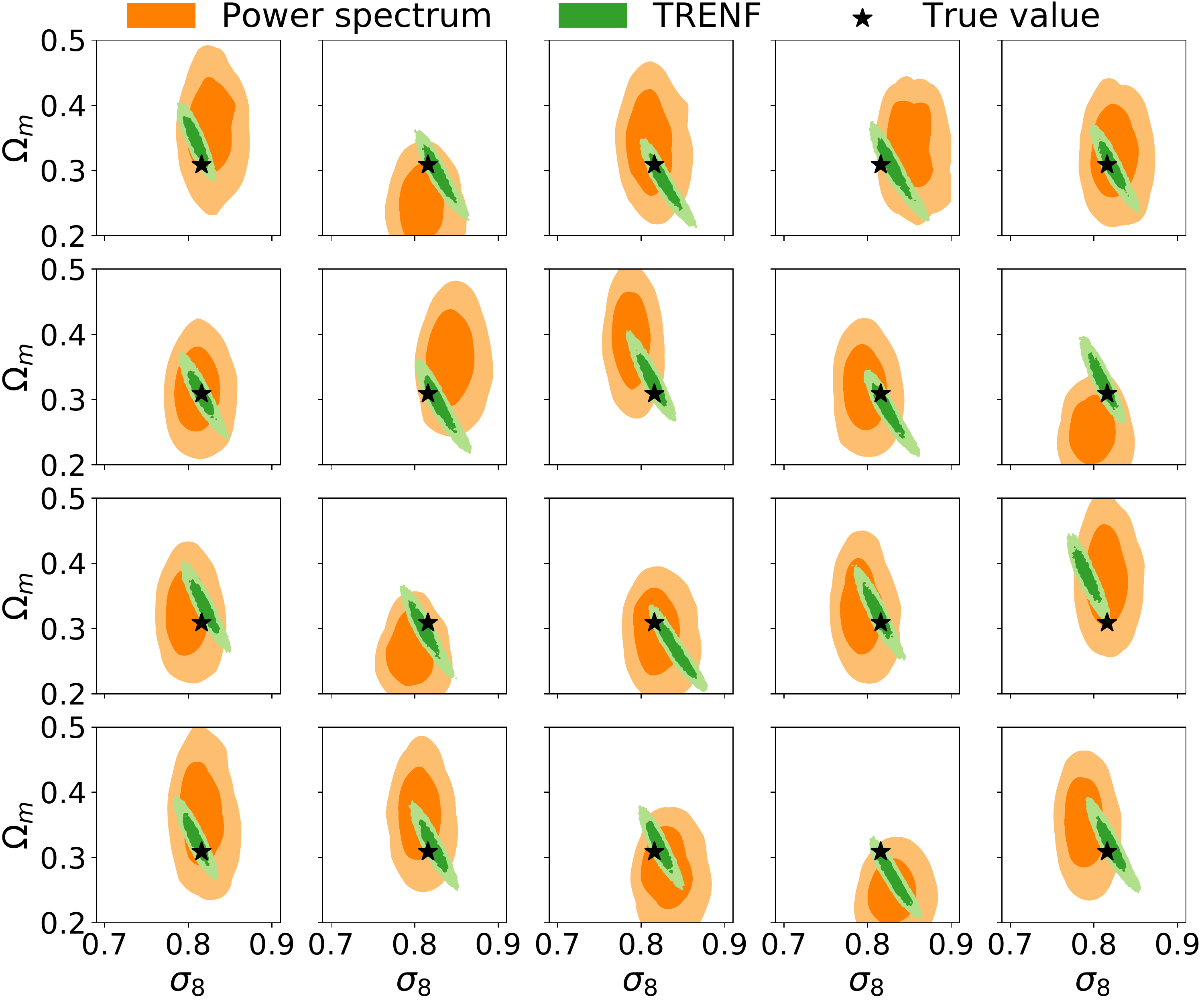}
    \caption{The posteriors from TRENF (green contour) and power spectrum (red contour) on uncurated test data. Figure of merit: 995 for TRENF, and 176 for power spectrum.}
    \label{fig:posterior}
\end{figure}

\begin{figure}
    \centering
    \includegraphics[width=\linewidth]{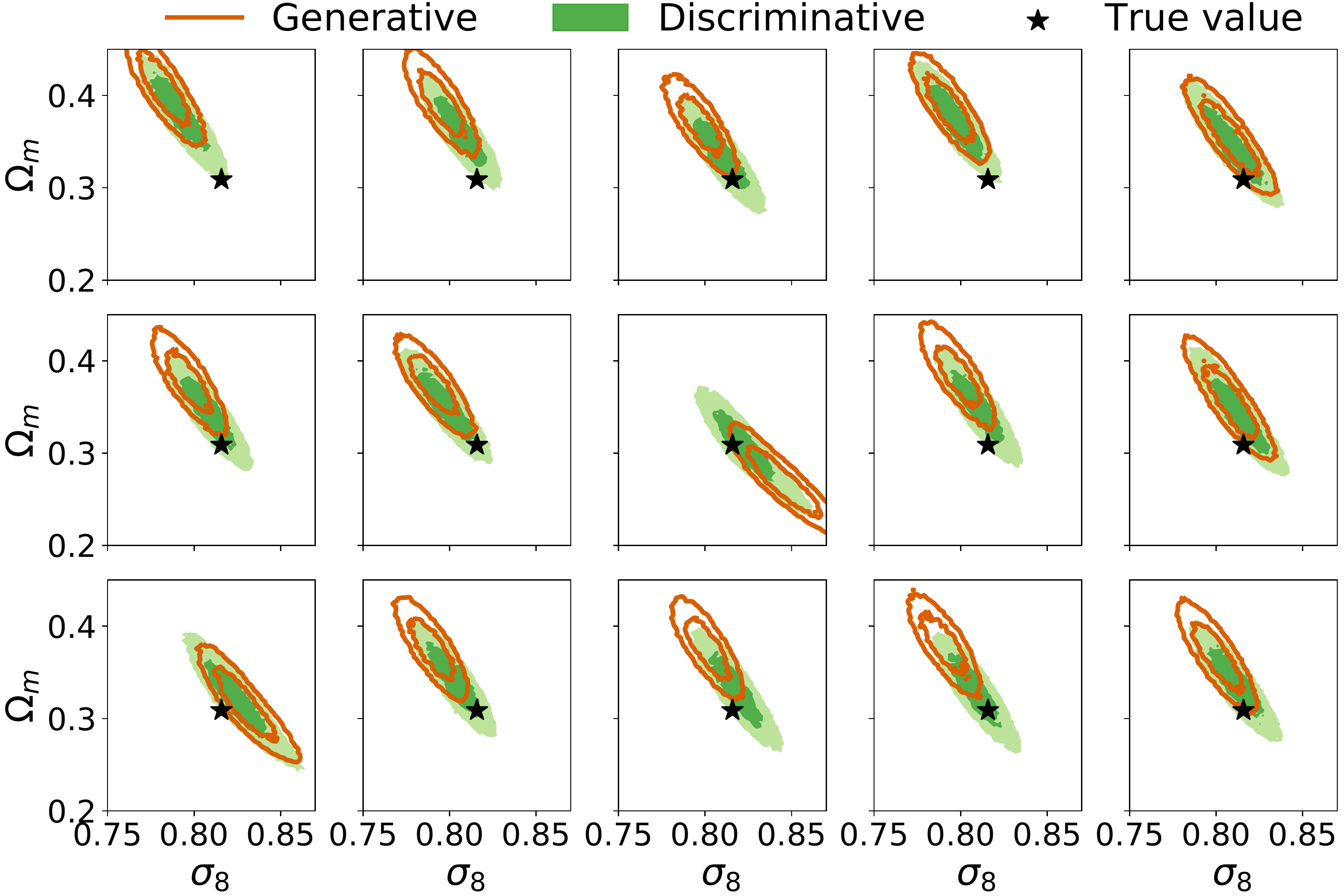}
    \caption{Comparison of the posterior constraints between generative learning (red contour) and discriminative learning (green contour) for the outlier cases in generative learning. Discriminative learning improves the posterior and results in properly calibrated posteriors, while generative learning is overconfident in the posteriors.}
    \label{fig:discriminative}
\end{figure}

We have shown that TRENF is able to learn the likelihood function accurately for GRFs. Now we explore the more challenging and more realistic dataset, the matter overdensity field. The dataset and the architecture of TRENF are the same as in Section \ref{sec:sample_latent}. We optimize TRENF using the two-stage training strategy as described in Section \ref{subsec:training} to improve the accuracy of posteriors.

In Figure \ref{fig:posterior} we present the $68\%$ and $95\%$ confidence regions of the posterior distribution on test data  (we assume Planck 2015 cosmology parameters). We compare the posterior constraints from TRENF with the standard power spectrum analysis. TRENF models the full likelihood function of the data vector $x$ without any dimension reduction, so it provides much tighter constraints than the power spectrum, which only uses two-point function information. We measure the figure of merit, defined as the inverse of the area of $68\%$ confidence region, on 100 test data for both methods. We obtain 995 for TRENF and 176 for power spectrum, which means that TRENF significantly improves the posteriors relative to the power spectrum.


On 100 test data, there are 65 cases where the true cosmology is within the $68\%$ confidence region, and 95 cases the true cosmology is within the $95\%$ region. These numbers are consistent with the $68\%$ and $95\%$ expectation, suggesting that the posteriors from TRENF are properly calibrated. Note that we need second-stage discriminative learning to achieve this. If we train TRENF only with the generative loss, we find the model is overconfident and we get more than 5\% outliers, where the true cosmology is outside the $95\%$ region, which is not consistent with the $5\%$ expectation. In Figure \ref{fig:discriminative} we show some outlier cases from the generative learning, and we see that discriminative learning (two stage learning) improves the posterior.

\section{Modeling the Data With Mask}
\label{sec:mask}

\begin{figure}
     \centering
     \begin{subfigure}[]{0.49\linewidth}
         \centering \includegraphics[width=\linewidth]{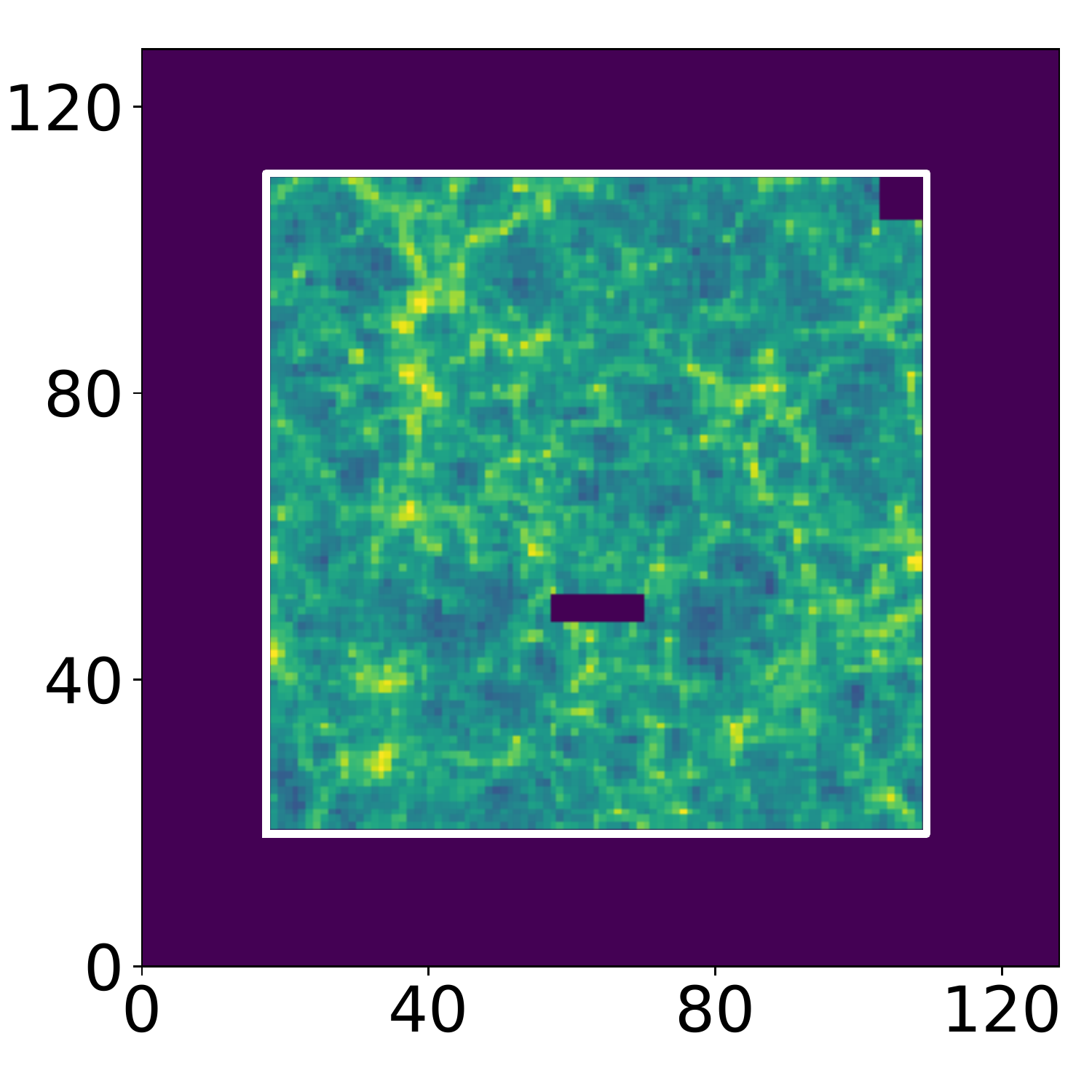}
     \end{subfigure}
     \hfill
     \begin{subfigure}[]{0.49\linewidth}
         \centering \includegraphics[width=\linewidth]{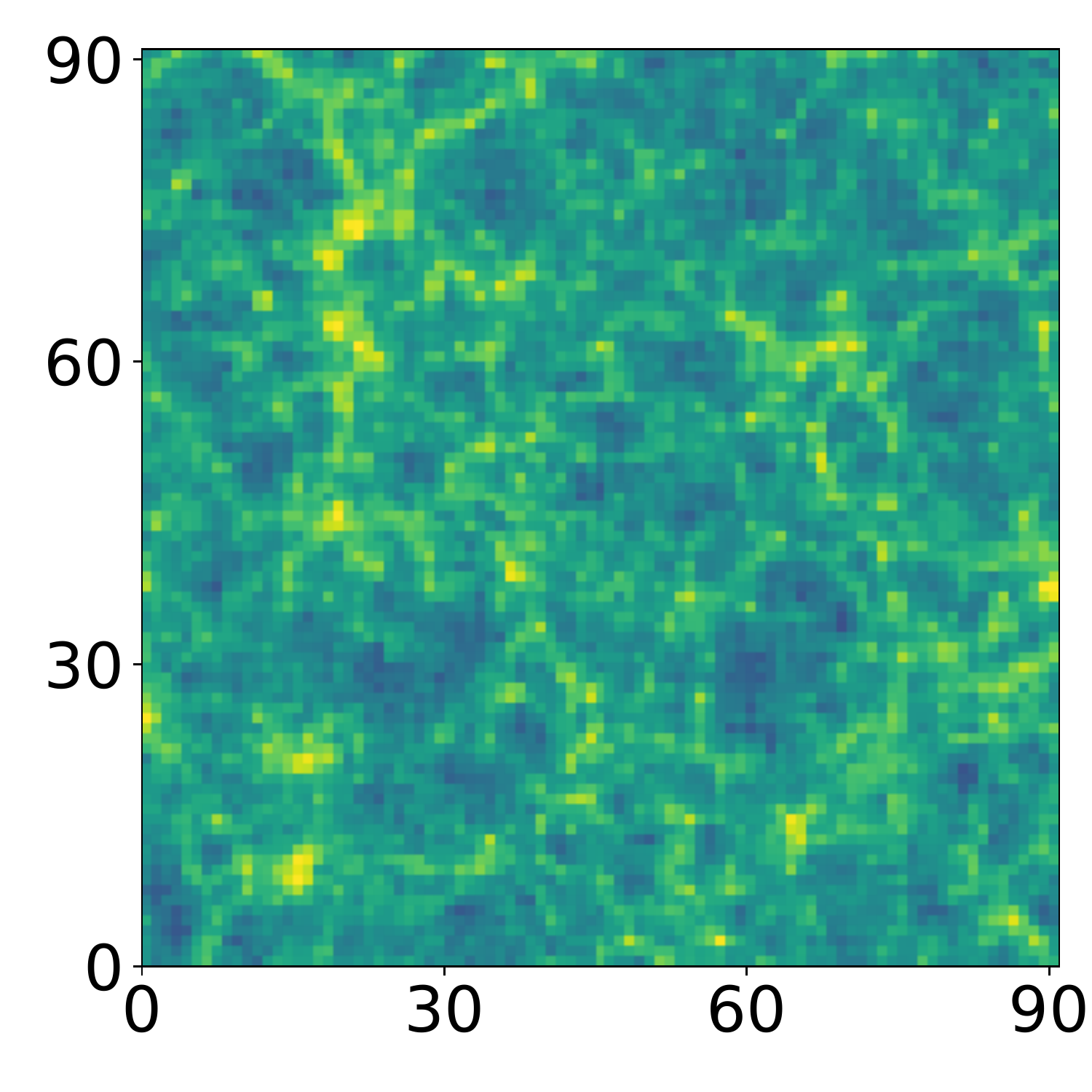}
     \end{subfigure}
    \caption{A visualization of the simplified mask we considered in this work (left panel), and the data after the affine coupling (inpainting) layer (right panel). In the left panel the blue regions represent missing pixels, and the white rectangle region denotes the data passed to the model.}
    \label{fig:mask}
\end{figure}

\begin{figure}
    \centering
    \includegraphics[width=\linewidth]{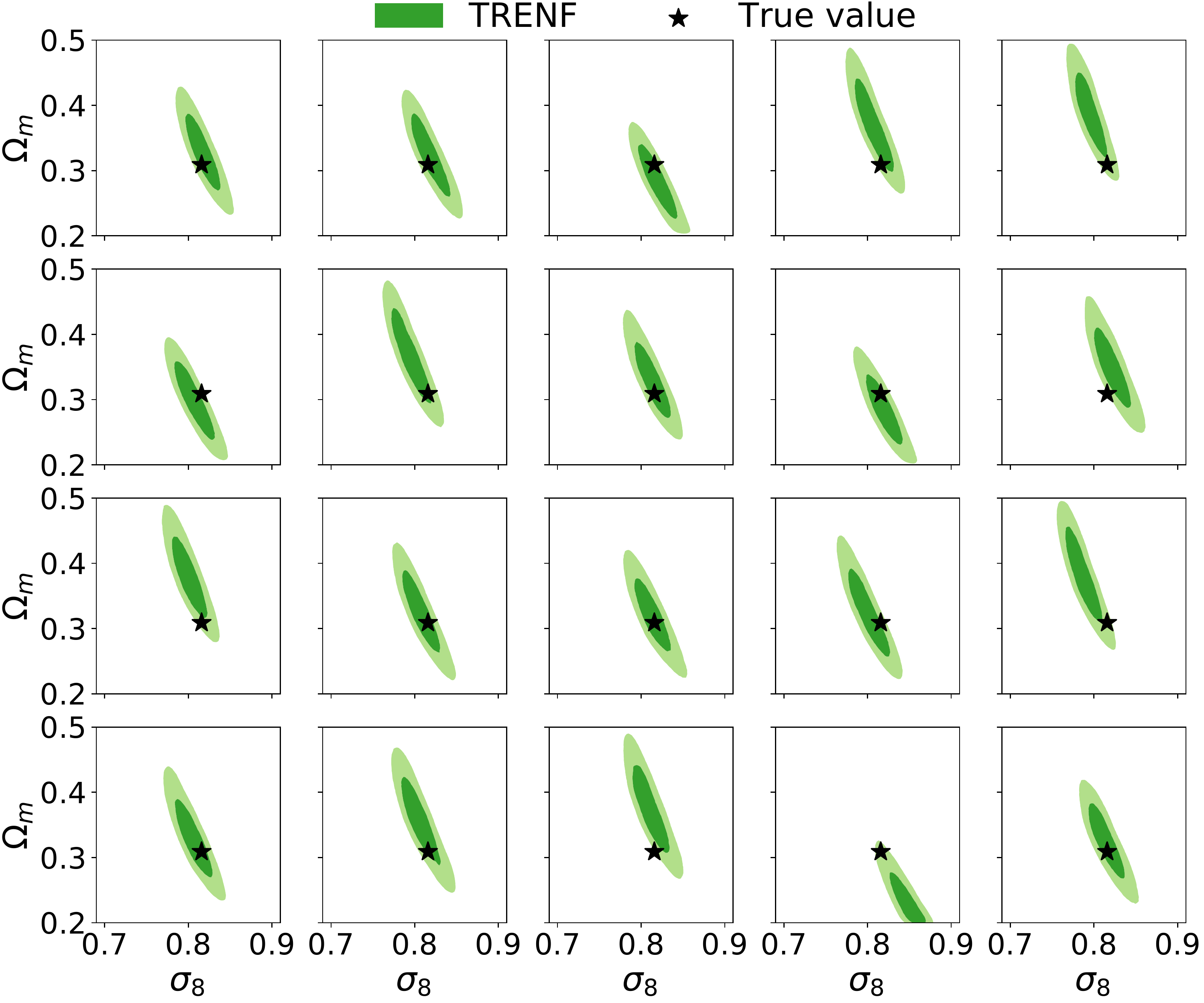}
    \caption{The posterior distribution from TRENF on test data with mask. $50\%$ of pixels is removed to mimic the survey mask. The figure of merit is 507.}
    \label{fig:posterior_mask}
\end{figure}

In this section we present a generalization of TRENF to model the effects that break the data symmetry. As described in Section \ref{subsec:mask}, our strategy is to first use an affine coupling layer to inpaint the missing pixels, and then introduce position dependence to the non-linearity $\Psi$ in Equation \ref{eq:TRENF} to model these effects. 

{\bf Dataset:}
We consider a simple example where we take the central $91\times 91$ pixels of the $128\times 128$ matter overdensity fields so that the boundaries are no longer periodic, and then remove the $6\times 6$ pixels at the upper right corner of the data to mimic the non-rectangular mask, and remove $4\times 13$ pixels in the center of the data to mimic the observational effects, such as foregrounds, cosmic ray hits, detector failures, etc. We show a visualization of this mask in the left panel of Figure \ref{fig:mask}. During training and inference, we sample Gaussian noise in the $6\times 6$ and $4\times 13$ missing pixels, and the full $91\times 91$ region is passed to the model for estimating the likelihood. Note that this is only a proof-of-principle study, and the mask we considered is a simplification to the realistic survey mask, but the methods we developed here should also apply to more realistic survey masks.

{\bf Model:}
Similar to Section \ref{sec:sample_latent}, we first apply an inverse softplus transform and a normalization transform on the observed pixels to remove the $[0,\infty)$ boundary and rescale. We then apply affine coupling transforms (Equation \ref{eq:affine}) on the Gaussian noise of the two missing regions. After the inpainting, we add 5 layers of convolutions and position-dependent non-linearities. Here we have two hyper networks $g_{\tilde{T}}$ and $g_{\Psi}$ (Equation \ref{eq:phi_T} and \ref{eq:phi_Psi}), and both of them, as well as the conditional networks in affine coupling layers, are chosen to be multilayer perceptrons with $2$ hidden layers and $512$ neurons in each hidden layer. The other hyperparameters and training strategies are the same as Section \ref{sec:posterior}.

We first show a visualization of the transformed data after inpainting in the right panel of Figure \ref{fig:mask}. Note that here we do not explicitly train the affine coupling layer to accurately recover the correct structures in the missing pixels. Instead, the goal of these layers is to inpaint structures that are statistically consistent with the observed data so that these missing pixels do not spoil the posterior analysis. To verify this, we show the posterior distribution of test data in Figure \ref{fig:posterior_mask}. We have verified that the uncertainty quantification is not miscalibrated: on 100 test data, there are 93 cases where the true cosmology is within the $95\%$ region, compared to the expected number of 95. If we assume that most of the information comes from the small scales, we expect that the amount of information would roughly be proportional to the area of the survey. Here we have removed about $50\%$ of the area as compared to the original dataset, so the constraining power should also be reduced by this amount. This is confirmed by our experiment: the figure of merit is now 507, as compared to 995 of the original dataset. This suggests that the amount of information extracted by TRENF is still close to optimal in the presence of the mask.

\section{Beyond spherical kernels}
\label{sec:DN}

As discussed at the end of Section \ref{subsec:TRENF}, TRENF can be viewed as a CNN with the number of channel $c=1$. In this section we will discuss TRENF in the framework of Steerable CNNs \citep{Cohen2017a,Weiler2018,Cohen2019}, which provides a general theory for equivariant networks. A steerable CNN defines the feature maps as steerable feature fields $v: \mathbb{R}^2 \rightarrow \mathbb{R}^c$. Under translation $t$ and rotation $r$, a steerable feature field $v(x)$ is transformed to $[\pi(tr)v](x)$, given by
\begin{equation}
\label{eq:steerable}
    [\pi(tr)v](x) = \rho(r) \cdot v(r^{-1}(x-t)) ,
\end{equation}
where $\rho(r)$ is the type of the feature field and is a representation of the symmetry group. For example, in TRENF we have $c=1$ and the feature field is a scalar field, which corresponds to the trivial representation $\rho(r)=1$. In general steerable CNNs one can also have vector fields, where $\rho(r)=r$ is the standard representation and Equation \ref{eq:steerable} becomes the familiar transformation law of vector fields. We refer the readers to \cite{Cohen2017a} and \cite{Cohen2019} for more details about steerable CNNs.

It has been shown that the most general linear map between steerable feature fields with type $\rho_{\mathrm{in}}$ and $\rho_{\mathrm{out}}$ is given by convolutions with kernel $T(x)$ satisfying \citep{Weiler2018,Cohen2019}
\begin{equation}
    T(rx) = \rho_{\mathrm{out}}(r) T(x) \rho_{\mathrm{in}}(r^{-1}) .
\end{equation}
In normalizing flows the transformation is invertible, so the dimensionality of the feature fields should stay the same between different layers and the representation $\rho_{\mathrm{in}}$ and $\rho_{\mathrm{out}}$ must all be one dimensional. For $\mathrm{O}(2)$ group (rotation and reflection) the only one dimensional real representation is the trivial representation $\rho=1$ \citep{Weiler2019a}, so we have $T(rx)=T(x)$ for any rotation $r$. Therefore spherical kernel is the only allowed kernel in TRENF.

\begin{figure}
     \centering
     \begin{subfigure}[]{0.49\linewidth}
         \centering \includegraphics[width=\linewidth]{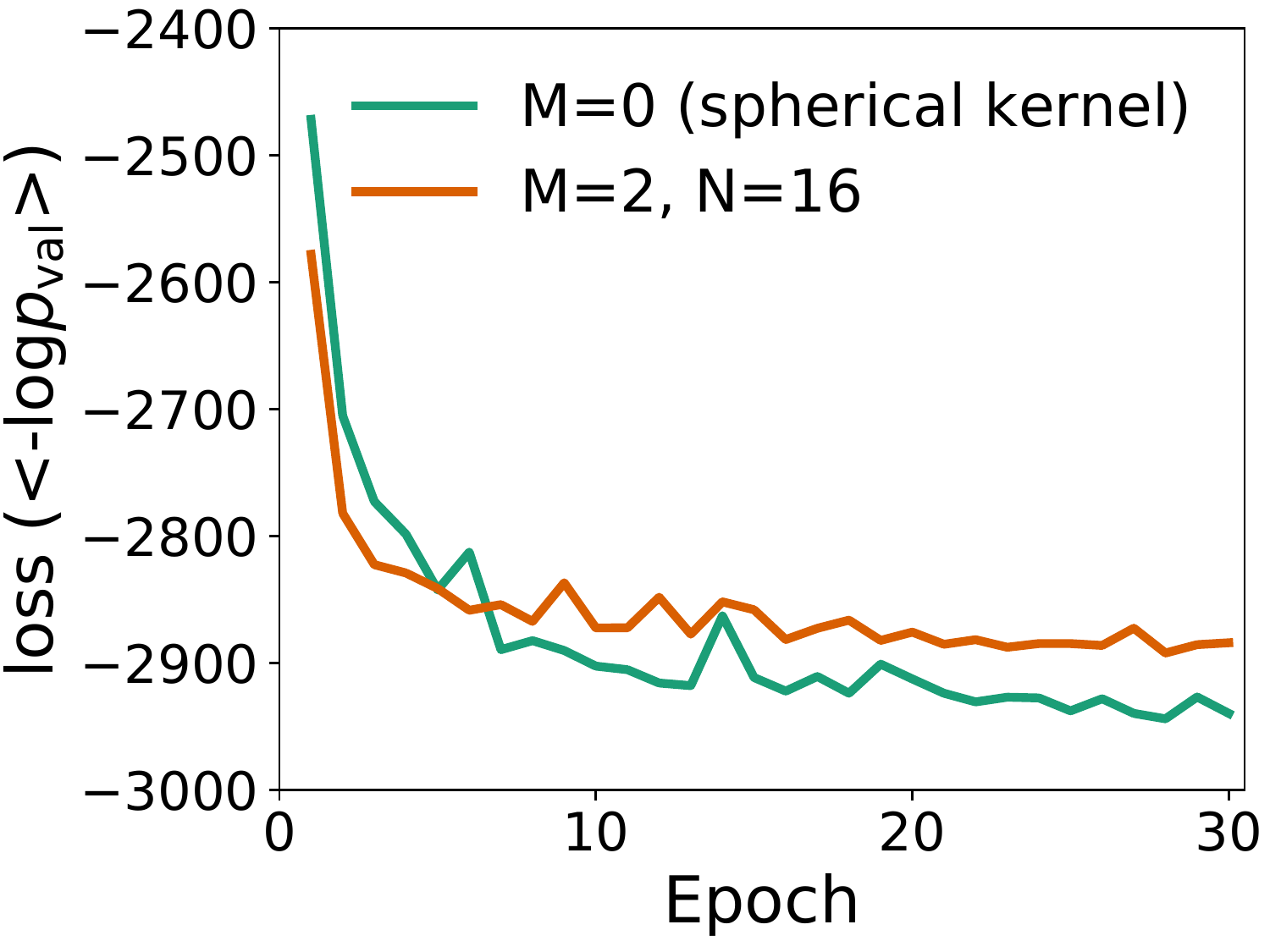}
     \end{subfigure}
     \hfill
     \begin{subfigure}[]{0.49\linewidth}
         \centering \includegraphics[width=\linewidth]{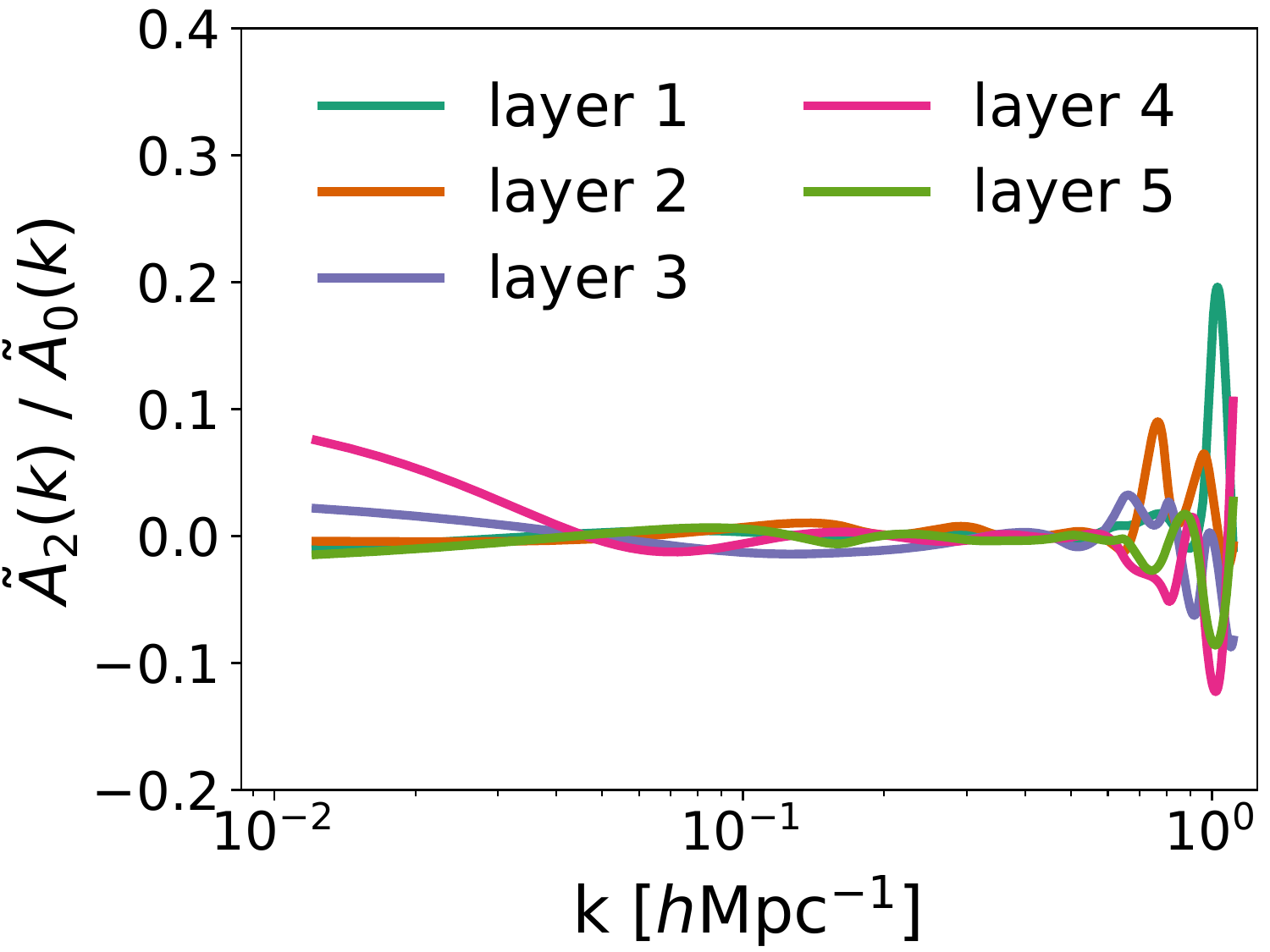}
     \end{subfigure}
    \caption{Left panel: the loss curve of TRENF with $M=0$ (spherical kernel) and $M=2, N=16$ on validation set. Right panel: the amplitude ratio $\tilde{A}_2(k)\ /\ \tilde{A}_0(k)$ of the convolution kernels in different layers of TRENF with $M=2, N=16$. See equation \ref{eq:DNkernel} for the definition of kernels we used in this experiment. TRENF with angular dependency $M=2$ performs worse than TRENF with spherical kernels, and the amplitude of non-isotropic terms is small compared to the spherical symmetric term $\tilde{A}_0$.}
    \label{fig:DN}
\end{figure}

To go beyond spherical kernels, one approach is to give up the exact $\mathrm{O}(2)$ symmetry and approximate it with $\mathrm{D}_N$ group (discrete rotations by angles multiple of $\frac{2\pi}{N}$ and reflection). With a sufficiently large $N$, $\frac{2\pi}{N} \rightarrow 0$ and its multiples can approximate any angles. $\mathrm{D}_N$ group has several 1D real representations. We refer the readers to Appendix F.2 and Table 12 of \cite{Weiler2019a} for irreducible representations of $\mathrm{D}_N$ group, as well as all possible convolutional kernels between different $\mathrm{D}_N$ representations. For simplicity and invertibility considerations, we explore convolution kernels of the following forms:
\begin{equation}
T(r,\phi) = A_0(r) + \sum_{t=1}^M A_t(r) \cos(tN\phi),
\end{equation}
where $A_0(r)$ and $A_t(r)$ are arbitrary radial functions. The above kernel ensures that the feature fields are all scalar fields and are equivariant under transformations of $\mathrm{D}_N$ group. In Fourier space, the kernel can be written as
\begin{equation}
\label{eq:DNkernel}
\tilde{T}(k,\theta) = \tilde{A}_0(k) + \sum_{t=1}^M (-i)^{tN}\tilde{A}_t(k) \cos(tN\theta),
\end{equation}
where $\theta$ is the polar angle of the Fourier $\bm{k}$ mode, and $\tilde{A}_t(k)$ is given by the Bessel function $J_{tN}$:
\begin{equation}
    \tilde{A}_t(k) = \int J_{tN}(kr)A_t(r)rdr.
\end{equation}
Similar to Section \ref{subsec:TRENF}, we directly parametrize $\tilde{A}_t(k)$ in Fourier space using cubic splines and they will be learned from the data. We replace the spherical TRENF kernel with Equation \ref{eq:DNkernel} and apply the model on the 2D matter overdensity fields as described in Section \ref{sec:sample_latent}. We tested different $N$ and $M$ choices ($N=\{8,16,32\}$, M=\{1,2,3\}), and compare their performance with the model with spherical kernels. For TRENF with high order terms $M>0$ we increase the width of the hyper network from 512 to 1024 for better conditional modeling, and keep the other hyperparameters the same. We find no improvements in terms of model loss (averaged negative data log-likelihood), sample quality, and posterior figure of merit. In Figure \ref{fig:DN} we show that TRENF with spherical kernels converges to a better loss than TRENF with $M=2, N=16$. We also show that high order term $\tilde{A}_2$ is relatively unimportant compared to the spherical symmetric term $\tilde{A}_0$. On small scales $\tilde{A}_2/\tilde{A}_0$ deviates from $0$, and this is probably because of imperfect optimization. $\tilde{A}_2/\tilde{A}_0=0$ is a better solution (left panel of Figure \ref{fig:DN}), but the model doesn't find it due to nonconvex optimization. This suggests that for the datasets we considered in this work, introducing extra angular dependence into the convolution kernel is not helpful and the spherical kernel is enough. 

Another way to combine steerable CNNs with NFs is to adopt affine coupling transforms \citep{dinh2016density} and use steerable CNN for the coupling network. The coupling network does not need to be invertible and one can use more complex kernels with multidimensional steerable feature fields. However, affine coupling transforms require splitting the data into two parts, and such splitting is not generally equivariant. This approach is beyond the TRENF architecture, and we will leave it for future studies.

\section{Discussion}

The main goal of this paper is to develop a Normalizing Flow with built in translation and rotation symmetry (TRENF), for the purpose of generating new samples and data likelihood analysis in cosmology. 
We argue that cosmological fields are high dimensional distributions which cannot be approximated by low dimensional manifolds, 
so low-dimensional-manifold models such as GANs or VAEs are not 
appropriate for this application. In addition, NFs give direct access to the data likelihood, which is of crucial significance for the data analysis.  
Because NFs preserve the dimensionality of the data, they usually do not scale well to high dimensions compared to other generative models such as GANs or VAEs, 
but here we argue that taking advantage of translation and rotation symmetries makes this task significantly easier. 

We argue that TRENF enables a clear 
path towards optimal cosmological analysis of the data, with a simple and computationally tractable approach. 
Specifically, we have shown that TRENF with only five layers saturates the information content that can be extracted from this architecture when applied to the 2D projections of cosmological N-body simulations, in the sense that adding more layers and using more complex kernels does not further improve the results.
While proving optimality of the transport map from one distribution to another is a notoriously difficult problem for high dimensional distributions, several lines of argument suggest our approach 
enables near optimal analysis for the 
application developed in this paper. 

First, when the method is applied to the Gaussian Random Fields with a known analytic solution, it extracts all the information correctly and optimally. 
Second, the inverse map from the data to the latent space is statistically 
indistinguishable from the target distribution of Gaussian white noise when we use correct cosmological parameters, while when the parameters are incorrect the map deviates from its target. 
This means that the non-Gaussian structures such as voids, filaments, and halos have all been mapped into Gaussian white noise, and all of their information has been transferred via the Jacobian of the transformation into $p(x|y)$. 
TRENF extracts all the information, including from the two-point correlations, one point distribution, and bispectrum, which can be observed from the fact that these statistics become non-informative once the data are mapped into the latent space. 
The main power of TRENF is that it is able not only to extract information from so many different statistics, but do it in a way that can be optimal, and for which the output is the data likelihood itself. 

TRENF can easily deal with noise in the data if noise is independent of position, as it preserves translation and rotation symmetry, and we can simply train TRENF on data with noise. 
A more difficult problem is that of cosmological data likelihood analysis in the presence of the survey mask. The mask breaks the translation and rotation symmetries, and is a notoriously difficult problem even for Gaussian fields: 
in the absence of the mask, the problem can be solved with Fast Fourier Transforms with $O(N\ln N)$ for flat geometry or $O(N^{3/2})$ for spherical geometry, while with the mask it requires a linear algebra solution that scales as $O(N^3)$, which becomes prohibitively expensive for large surveys. 
We introduce position dependence to the pointwise nonlinearity at each layer as the non-equivariant component in our model. We show that this approach is fast to train and to evaluate the likelihood. The constraining power (figure of merit) is reduced by the fraction of the area of the mask, consistent with our expectations if we assume most of the information comes from the small scales. This suggests that our approach is still optimal when the survey mask is included.
Note that this method can also model other processes that break translation and rotation symmetry, for example, position-dependent noise, seeing, foregrounds, etc. 

Potential applications and future generalizations of TRENF are numerous, here we list a few examples:
\begin{itemize}
\item TRENF enables the possibility of fast training and generation of new cosmological data outputs from a few existing simulations. 
This has numerous applications such as Lyman alpha forest, 21cm, and other intensity maps, weak lensing maps, projected galaxy clustering, X-ray and thermal SZ maps, etc. 
We expect that TRENF can learn efficiently with fewer training input maps (less data complexity), as compared to previous generative approaches, a consequence of translation and rotation symmetry built into the model. TRENF should also be generalized to output multiple 
maps of different tracers on the same area of the sky. 
\item TRENF training of latent space does not directly impose a spatial structure, although in practice we observe a strong correlation 
between the latent space and data space. It may be possible to make that more explicit, by enforcing the 
latent space to be the initial conditions of a simulation. In addition, one can also train TRENF as a function of time. In this case, TRENF would become an Eulerian N-body or hydro simulation. TRENF can also be used to learn the velocity field, to 
describe the full phase space information. 
\item TRENF has the ability to perform nearly optimal posterior inference analysis of cosmological parameters via the data likelihood, given the TRENF's ability to evaluate $p(x|y)$. 
We emphasize that TRENF learns the data likelihood directly, and there is no need to learn the probability distribution of the summary statistics.
One can view TRENF as a way to optimally combine all of the summary statistics proposed in the cosmology literature, such as two-point, three-point, and higher order correlations, one-point 
distribution of various smoothing scales, 
void profile and void-void correlations, 
void and halo mass functions, topological 
statistics, etc. To the extent that the data 
have been mapped to Gaussian white noise, all 
of these summary statistics have been used optimally, by extracting their contribution 
to the Jacobian in $p(x|y)$. 
\item 
In addition to cosmological parameters, one 
can also train TRENF on astrophysical parameters, such as baryonic processes. These can be incorporated into the full likelihood analysis, by training TRENF on hydrodynamic simulations \citep{Villaescusa2021a} or baryon maps generated by fast machine learning \citep{Dai2021a, Dai2018a}, or semi-analytical approaches \citep{Aric02020a}. Once we have the likelihood of the data as a function of these parameters we can marginalize over these effects in the posterior analysis.

\item TRENF can be generalized to 3D galaxy redshift space data, where observed redshift is the sum of position and velocity of the galaxy, and we no longer have exact rotation symmetry. 
Instead, we must describe the data in terms of the line of sight and perpendicular to the line of sight coordinates or their harmonic
transforms, similar to our 
2D expansion in equation \ref{eq:DNkernel}. 
\item TRENF can be used to search for 
primordial non-Gaussianity in the latent space. Because the non-Gaussianity from the 
nonlinear evolution of structure is eliminated
in the latent space, it becomes easier to 
search for other non-Gaussian effects, such as 
primordial non-Gaussianity. 
\end{itemize}


\section*{Acknowledgements}
This material is based upon work supported by the National Science Foundation under Grant Numbers 1814370 and NSF 1839217, by NASA under Grant Number 80NSSC18K1274, and by the U.S. Department of Energy, Office of Science, Office of Advanced Scientific Computing Research under Contract No. DE-AC02-05CH11231 at Lawrence Berkeley National Laboratory to enable research for Data-intensive Machine Learning and Analysis.

\section*{Data Availability}

The code and simulation data generated in this research will be shared on reasonable request to the corresponding author.



\bibliographystyle{mnras}
\bibliography{reference} 

\bsp	
\label{lastpage}
\end{document}